\documentclass[pdflatex,tex,twocolumn,epjc3]{svjour3}          

\RequirePackage[T1]{fontenc}

\smartqed  
\usepackage{amsmath}
\usepackage{indentfirst}
\usepackage{amssymb,dcolumn}
\usepackage{latexsym}
\usepackage{graphics}
\usepackage{setspace}
\usepackage{graphicx,float}
\usepackage[colorlinks,linkcolor=blue,citecolor=blue,urlcolor=blue,hyperindex]{hyperref}
\usepackage{color}
\usepackage{slashed}
\usepackage{CJKutf8}
\RequirePackage{graphicx}
\RequirePackage{mathptmx}      
\RequirePackage{flushend}
\RequirePackage[numbers,sort&compress]{natbib}
\RequirePackage[colorlinks,citecolor=blue,urlcolor=blue,linkcolor=blue]{hyperref}

\journalname{Eur. Phys. J. C}

\begin{document}
\begin{CJK}{UTF8}{<font>}
\title{Optical appearance of a thin-shell wormhole with a Hayward profile}

\author{Sen Guo\thanksref{addr1,e1}
        \and
        Guan-Ru Li\thanksref{addr1,e2}
        \and
        En-Wei Liang\thanksref{addr1,e3}}
\thankstext{e1}{e-mail: sguophys@126.com}
\thankstext{e2}{2007301068@st.gxu.edu.cn}
\thankstext{e3}{Corresponding author, e-mail: lew@gxu.edu.cn}
\institute{\label{addr1} Guangxi Key Laboratory for Relativistic Astrophysics, School of Physical Science and Technology, Guangxi University, Nanning 530004, People's Republic of China}

\date{Received: date / Accepted: date}

\maketitle

\begin{abstract}
The optical properties of a thin-shell wormhole (TSW) with a Hayward profile is investigated. Adopting the ray-tracing method, we demonstrate that the TSW's contralateral spacetime is capable of reflecting a significant portion of light back to the observer spacetime. We analyze the effective potential, light deflection, and azimuthal angle of the TSW and find that these quantities are affected by the mass ratio of the black holes (BHs). Specifically, if the mass of the contralateral spacetime BH is greater than that of the original spacetime BH, and the impact parameter satisfies the condition $Hb_{\rm c2}<b_{1}<b_{\rm c1}$, the trajectory of the photon exhibits round-trip characteristics. Assuming the presence of a thin accretion disk surrounding the observing spacetime BH, our results indicate that the image formed by the TSW exhibits additional photon rings and a lensing band compared to an image produced by a BH alone.
\end{abstract}

\section{Introduction}
\label{sec:intro}
\setlength{\parindent}{2em}
Black holes (BHs) are intriguing astronomical objects whose existence is predicted by the theory of general relativity (GR). One of their most captivating characteristics is their ability to cause gravitational lensing of light. Since photons follow the null geodesics of the spacetime metric, studying the geodesic equation provides valuable insights into the bending of light. In an attempt to describe the deflection of light around a BH using the concept of a BH shadow, Synge determined that the shape of a spherical BH (Schwarzschild BH) shadow is a perfect circle \cite{1}. Building upon the Schwarzschild BH model, Bardeen proposed that the deformation of the BH shadow is related to the BH's angular momentum \cite{2}, specifically in the context of rotating BHs (Kerr BHs). Extensive research has been conducted to investigate the geometry of BH shadows in various modified gravity contexts were also discussed in the existing literatures \cite{3,4,5,6,7,8,9,10,11,12,13,14,15,16,17,18,19,20,21,22,23,24}.
\\
\indent
In astrophysical observations, it is common to find a luminous accretion disk surrounding a BH. Luminet demonstrated that the presence of the BH shadow and ring depends on the position of the accretion disk \cite{25}. Investigating a scenario with a spherical accretion disk, Narayan $et~al.$ studied the optical appearance of a Schwarzschild BH and showed that the size of the BH shadow remains unaffected by the radius of the spherical accretion disk \cite{26}. When considering a thin accretion disk surrounding the Schwarzschild BH, Gralla $et~al.$ presented an elegant description of the BH shadow, lensing ring, and photon ring \cite{27}. They discovered that the brightness of the shadow exhibits a logarithmic divergence at the photon ring and that the details of the accretion affect the BH shadow. The impact of the location and morphology of the accretion flow surrounding the BH spacetime has been extensively investigated in various modified gravity theories \cite{28,29,30,31,32,33,34,35,36,37,38}.
\\
\indent
By employing numerical simulations and ray-tracing techniques, Falcke $et~al.$ demonstrated the feasibility of observing the BH shadow \cite{39}. The Event Horizon Telescope (EHT) collaboration achieved a breakthrough by obtaining the first-ever image of the supermassive BH located at the center of the elliptical galaxy Messier 87$^{*}$ (M87$^{*}$) \cite{40,41,42,43,44,45}. This image revealed a prominent ring-shaped structure surrounding the BH shadow. EHT recently captured horizon-scale radio observations of Sagittarius A$^{*}$ (Sgr A$^{*}$), the supermassive BH in our own Milky Way. The measured size of the ring in Sgr A$^{*}$ aligns with the predicted critical curve of the BH shadow within a 10$\%$ margin of error, providing compelling evidence for GR in the regime of strong gravitational fields \cite{46,47,48,49,50,51}. These remarkable findings underscore the valuable insights the BH shadow offers regarding the spacetime properties near the BH, establishing it as a direct and powerful experimental confirmation of GR in the realm of strong gravity \cite{52,53,54}.
\\
\indent
The groundbreaking results obtained by the EHT have sparked extensive research into the optical properties of non-Kerr objects, which include BHs that deviate from the Kerr metric and horizonless ultra-compact objects \cite{55,56}. This research explores the characteristics of shadows cast by specific types of compact objects, such as extended Kerr BHs, additional scalar fields \cite{57,58}, horizonless objects like boson stars \cite{59}, and wormholes \cite{60,61,62}. A thin-shell wormhole (TSW) is a theoretical construct in the framework of GR that represents a ``shortcut'' through spacetime, connecting two distant regions. It consists of an exceedingly thin spherical shell of matter, enabling particles to pass through it and enter the ``throat'' of the wormhole. The shell serves as a sort of ``tunnel'' that allows particles to traverse vast distances in the universe without having to travel through regular space. TSWs serve as important models resembling BHs. With the exception of the innermost part, their spacetime can be identical to that anywhere in a BH. Visser proposed an effective method for describing and constructing a class of wormholes using the cut-and-paste technique \cite{63}, which involves combining two BH spacetimes. Tsukamoto investigated the gravitational lensing effect of the Schwarzschild TSW and discovered that the formation of images depends on the positions of the light source and the observer \cite{64}. Wang $et~al.$ explored the influence of the unstable photon sphere on TSW and found that the TSW's shadow size is always smaller than that of a BH \cite{65}. Researchers examining an asymmetric Schwarzschild TSW accretion disk demonstrated that the structure of the TSW's shadow differs from that of a BH \cite{66}. Guerrero $et~al.$ argued that wormhole geometries and proposed the existence of additional light rings in the intermediate region between the critical curves of the TSW \cite{67}. More recently, the photon-ring structure of horizonless objects characterized by two or no photon spheres has been explored \cite{68}. The existence of multi-ring images with a non-negligible luminosity in shadow observations when one allows for the existence of other compact objects \cite{69,70,aa,bb,cc}.
\\
\indent
In contrast to irregular BHs that possess an inherent singularity at the origin of spacetime, Bardeen proposed a BH solution that is devoid of spacetime singularities \cite{71}. Additionally, Hayward developed the regular Hayward BH solution by combining nonlinear electrodynamics with the Einstein field equations \cite{72}. We investigated the effects of accretion flow and magnetic charge on the observable characteristics of the shadow and photon rings of the Hayward BH \cite{73}. It is found that the geometric appearance of the Hayward BH shadow is determined by the underlying spacetime geometry, whereas the optical appearance is influenced by the properties of the accretion flow and the BH's magnetic charge.
\\
\indent
In this analysis, we utilize the cut-and-paste technique proposed by Visser \cite{63} to construct a TSW with a Hayward profile and investigate the optical properties of this particular class of horizonless objects. It should be noted that the construction of such a TSW involves connecting two spacetime structures with Hayward profiles. The structure of this work is organized as follows. In Section \ref{sec:2}, we provide a brief introduction to the construction of a TSW with a Hayward profile and discuss the effective potential. Section \ref{sec:3} presents the analysis of light deflection in the TSW system. In Section \ref{sec:4}, we delve into the classification of rings, transfer functions, and the corresponding optical appearance when a thin accretion disk surrounds the TSW. We draw the conclusions in Section \ref{sec:5}.

\section{\textbf{Effective potential of a TSW with a Hayward profile}}
\label{sec:2}
\par
\begin{figure}[tbp]
\centering
\includegraphics[width=7cm,height=5cm]{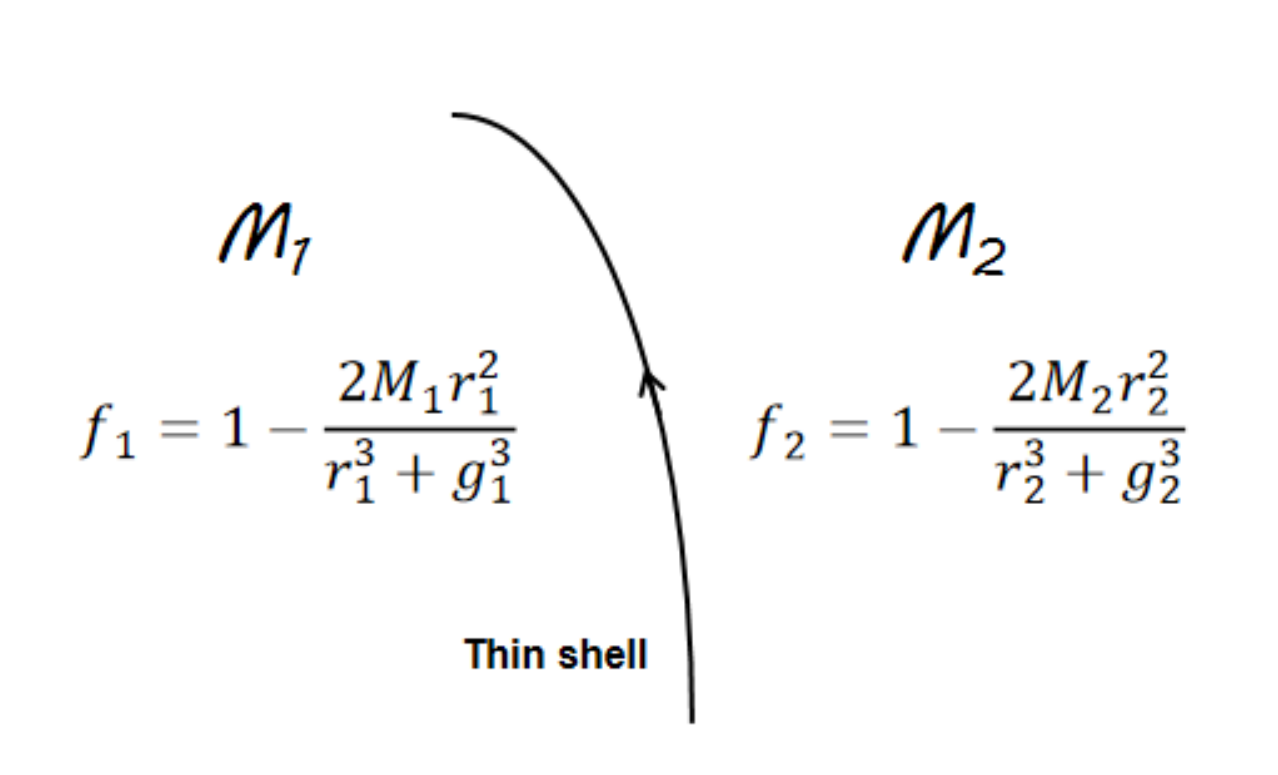}
\caption{\label{fig:1} A TSW with a Hayward profile serves as a mimic of a BH. The throat of the TSW establishes a connection between two distinct regions in spacetime, allowing for a light ray to pass through and link the two regions without crossing the event horizon.}
\end{figure}

The Hayward BH metric can be written as \cite{72}
\begin{equation}
\label{2-2}
{\rm d}s^{2}=-f(r) {\rm d}t^{2} + f(r)^{-1}{\rm d}r^{2} + r^{2}({\rm d} \theta^{2} + \sin^{2} \theta {\rm d} \phi^{2}),
\end{equation}
where $f(r)$ is the metric potential,
\begin{equation}
\label{2-3}
f(r)=1-\frac{2 M r^{2}}{r^{3}+g^{3}}.
\end{equation}
The symbol $M$ represents the mass of the BH, and $g$ denotes the BH magnetic charge. To create a TSW with a Hayward profile, we remove the interior regions of two Hayward spacetimes and join them together at the hypersurface. The two original spacetimes are denoted as $\mathcal{M}_{1}$ and $\mathcal{M}_{2}$, and the resulting manifold can be represented as $\mathcal{M} \equiv \mathcal{M}_{1} \cup \mathcal{M}_{2}$. The connection between the two spacetimes forms a throat, as illustrated in Fig. \ref{fig:1}.

\par
Based on Eq. (\ref{2-2}), the two spherically symmetric metrics are connected by the throat, we have
\begin{equation}
\label{2-4}
{\rm d}s_{\rm i}^{2}=-f_{\rm i}(r_{\rm i}){\rm d}t_{\rm i}^{2}+f_{\rm i}(r_{\rm i})^{-1}{\rm d}r_{\rm i}^{2}+r_{\rm i}^{2}({\rm d} \theta_{\rm i}^{2}+\sin^{2}\theta_{\rm i} {\rm d} \phi_{\rm i}^{2}),
\end{equation}
where $i=1,~2$ represent two spacetimes $\mathcal{M}_{1}$ and $\mathcal{M}_{2}$. The $f_{\rm i}(r_{\rm i})$ is
\begin{equation}
\label{2-5}
f_{\rm i}(r_{\rm i})=1-\frac{2 M_{\rm i}r_{\rm i}^{2}}{r_{\rm i}^{3}+g_{\rm i}^{3}},~~~~~~~~r \geq R,
\end{equation}
where $R$ denote the throat radius, satisfying $R > \max\{r_{+_{1}},r_{+_{2}}\}$ ($r_{+}$ is the event horizon radius). When the magnetic charge is zero in both spacetimes, the throat connects two Schwarzschild spacetimes. If the magnetic charge is zero in one of the spacetime, the throat connects a Schwarzschild spacetime with a Hayward spacetime. If the magnetic is non-zero charge in both spacetimes, the throat connects two Hayward spacetimes.

\par
To investigate the deflection of light in a TSW with a Hayward profile, we derive the effective potential governing its motion. We focus on photons that move exclusively on the equatorial plane. Assuming that the interaction between the photon and the throat is solely influenced by gravity, the 4-momentum of the photon remains invariant when crossing the throat. Considering the TSW definition, which ensures the continuity of the metric in the spacetime $\mathcal{M}$, i.e., $g_{\mu\nu}^{\mathcal{M}{1}}(R) = g_{\mu\nu}^{\mathcal{M}_{2}}(R)$ \cite{65}, we can express the motion equation of the null geodesic as follows:
\begin{equation}
\label{2-6}
\frac{p_{\rm t_{\rm i}}^{2}}{f_{\rm i}(r_{\rm i})}-\frac{p_{\rm \phi_{\rm i}}^{2}}{r_{\rm i}^{2}}=\frac{(p_{\rm i}^{r_{\rm i}})^{2}}{f_{\rm i}(r_{\rm i})},
\end{equation}
where $p^{\rm \mu}={{\rm d}x^{\rm \mu}}/{{\rm d} \lambda}$ represents the 4-momentum of the photon, $\lambda$ is an affine parameter. $p_{\rm t_{\rm i}}$ and $p_{\rm \phi_{\rm i}}$ denote the energy ($p_{\rm t_{\rm i}}=-E_{\rm i}$) and angular momentum ($p_{\rm \phi_{\rm i}}=L_{\rm i}$) of the photon, respectively, which remain constant along the geodesic due to the Killing symmetries of spacetime. By utilizing Eq. (\ref{2-6}), we can express the radial component of the null geodesic as follow:
\begin{equation}
\label{2-7}
p_{\rm i}^{r_{\rm i}} = \pm E_{\rm i} \sqrt{1 - \frac{b_{\rm i}^{2}}{r_{\rm i}^{2}}f_{\rm i}(r_{\rm i})},
\end{equation}
where $b$ is the impact parameter, defining as $b \equiv p_{\rm \phi_{\rm i}} / p_{\rm t_{\rm i}} = |L| / E$. From Eq. (\ref{2-7}), the effective potential $V_{\rm eff}$ of a TSW with a Hayward profile is obtained, i.e.,
\begin{equation}
\label{2-8}
V_{\rm eff_{\rm i}}= \frac{f_{\rm i}(r_{\rm i})}{r_{\rm i}^{2}} = \frac{1}{r_{\rm i}^{2}} \Bigg(1-\frac{2 M_{\rm i}r_{\rm i}^{2}}{r_{\rm i}^{3}+g_{\rm i}^{3}}\Bigg).
\end{equation}
The photon ring orbit satisfies the effective potential critical conditions
\begin{equation}
\label{2-9}
V_{\rm eff_{\rm i}}(r_{\rm ph_{\rm i}})=\frac{1}{b_{\rm c_{\rm i}}^{2}},~~~~~V'_{\rm eff_{\rm i}}(r_{\rm ph_{\rm i}})=0,
\end{equation}
in which $r_{\rm ph}$ denotes the radius of the photon ring, and $b_{\rm c}$ represents the critical impact parameter. Table. \ref{tab:1} presents the values of critical impact parameters for a TSW with a Hayward profile at various magnetic charges. The findings reveal that as the magnetic charge ($g$) increases, the critical impact parameter ($b_{\rm c_{i}}$) decreases. This indicates that the presence of a higher magnetic charge causes the photon ring to be pushed inward towards the BH, resulting in a smaller critical impact parameter.
\begin{table}[tbp]
\centering
\begin{tabular}{|l|r|c|c|c|c|c|c|c}
\hline
  $b_{\rm c_{i}}$/$g$ & $0~~~~$ & $0.2$ & $0.4$ & $0.6$ & $0.8$ \\
  \hline
  $~~b_{\rm c_{1}}$ & $5.19615$ & $5.19461$ & $5.18373$ & $5.15336$ & $5.09013$ \\
  \hline
  $~~b_{\rm c_{2}}$ & $6.23538$ & $6.23431$ & $6.22679$ & $6.20603$ & $6.16412$\\
  \hline
\end{tabular}
\caption{\label{tab:1} The critical impact parameters for a TSW with a Hayward profile, considering different magnetic charges. The BH masses are set as $M_{1}=1$ and $M_{2}=1.2$, while the magnetic charge is varied as $g=0,~0.2,~0.4,~0.6,~0.8$.}
\end{table}

\par
\begin{figure}[tbp]
\centering
\includegraphics[width=8cm,height=5cm]{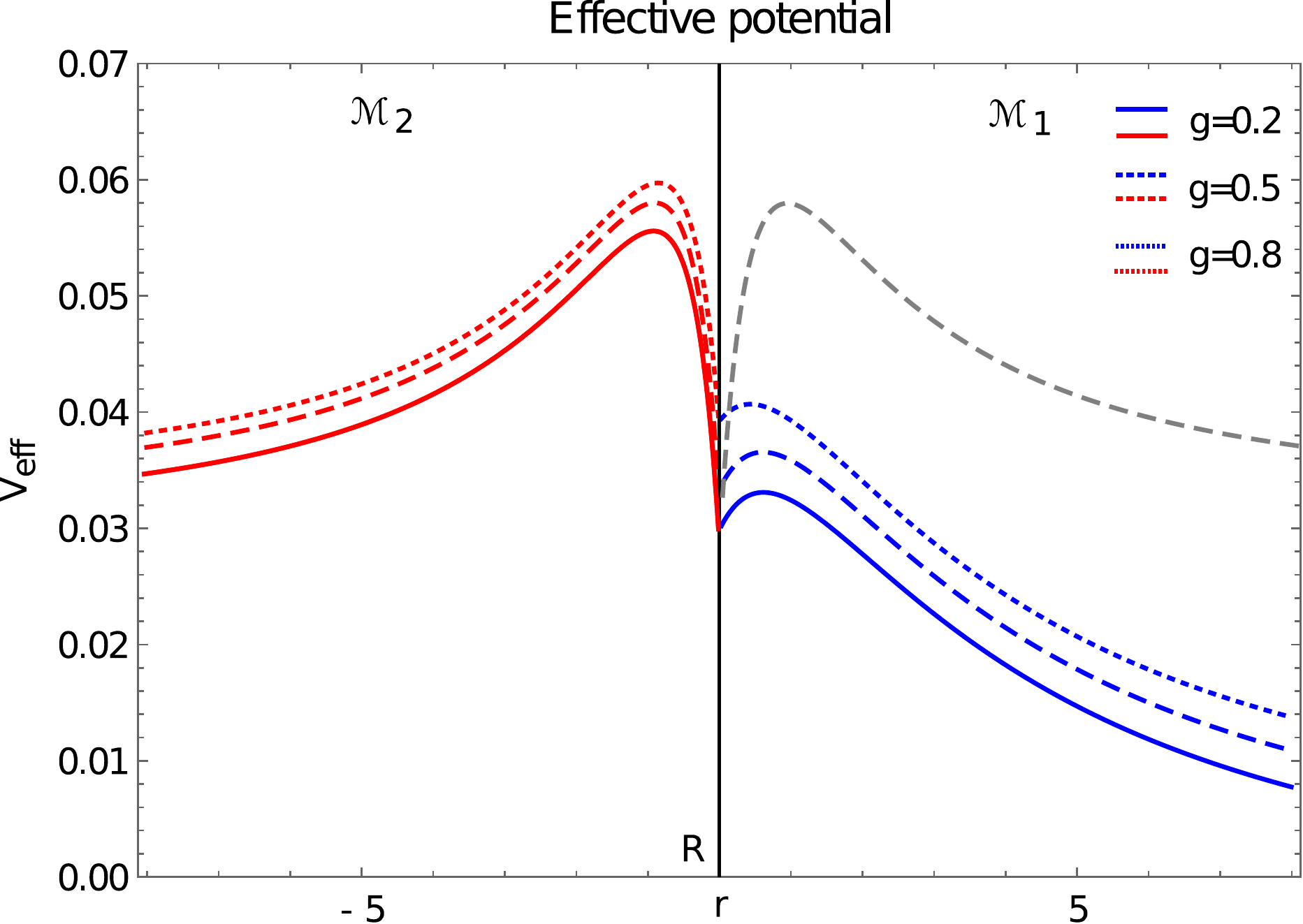}
\caption{\label{fig:2} The effective potential of a TSW with a Hayward profile as a function of radius for different magnetic charge. The blue and red curves two Hayward spacetimes with BH masses $M_{1}=1$ and $M_{2}=1.2$, respectively. The throat radius is chosen as $R=2.6$. The gray curve represents the effective potential function with the same spacetime parameters, namely $M_{1}=M_{2}=1.2$ and $g_{1}=g_{2}=0.5$.}
\end{figure}

Figure \ref{fig:2} depicts the effective potential of a TSW with a Hayward profile as a function of radius, revealing notable difference compared to the case of a Hayward BH. The throat of the TSW connects two distinct spacetimes, denoted as $\mathcal{M}_{1}$ and $\mathcal{M}_{2}$. When the masses satisfy $M_{1}=M_{2}$, the effective potential functions are identical on both sides of the throat, indicating consistency in the critical curves of the two spacetimes. However, if $M_{1}<M_{2}$, although the effective potential is equal at the throat, the highest point of the effective potential curve in spacetime $\mathcal{M}_{2}$ exceeds that in $\mathcal{M}_{1}$. This implies that the effective potential in spacetime $\mathcal{M}_{2}$ can reflect some light from spacetime $\mathcal{M}_{1}$. Furthermore, we observe that a larger magnetic charge leads to a more pronounced peak in the effective potential, which is consistent with the behavior observed in the case of the Hayward BH.

\section{\textbf{Light trajectory and deflection angle of a TSW with a Hayward profile}}
\label{sec:3}
\par
The phenomenon of light deflection in a BH can be categorized into three distinct situations. Firstly, when the impact parameter ($b$) is greater than the critical impact parameter ($b_{\rm c}$), the gravitational force causes the light ray to be deflected towards the observer. Secondly, when $b$ is smaller than $b_{\rm c}$, the light ray enters the BH and ultimately falls into the singularity. Lastly, when $b$ is equal to $b_{\rm c}$, the light ray orbits multiple times along the photon orbit of the BH, resulting in the formation of a luminous ring \cite{27}. However, the behavior of photons in a TSW differs significantly from that in a BH. Consequently, a reevaluation of light deflection is necessary in the analysis of TSW.

\par
Let's assume that light enters the spacetime $\mathcal{M}_{1}$ from infinity with an impact parameter of $b_{1}$. The deflection of light can be classified into three different scenarios depending on the mass relationship between the two spacetimes:
\begin{itemize}
\item \emph{Case 1}~~$M_{1}=M_{2}$: The critical impact parameter satisfies $b_{\rm c_{1}} = b_{\rm c_{2}}$. The light passes through the throat into the spacetime $\mathcal{M}_{2}$ within the necessary condition $b_{1} < b_{\rm c_{1}}$, and then returns to the spacetime $\mathcal{M}_{1}$.
\item \emph{Case 2}~~$M_{1}>M_{2}$: The critical impact parameter satisfies $b_{\rm c_{1}} > b_{\rm c_{2}}$. If $b_{1} < b_{\rm c_{2}}$, the light passes through the throat into the spacetime $\mathcal{M}_{2}$, and then returns to the spacetime $\mathcal{M}_{1}$. If $b_{1} = b_{\rm c_{2}}$, the light will provide additional brightness for the BH photon ring in the spacetime $\mathcal{M}_{2}$. If $b_{1} > b_{\rm c_{2}}$, the light goes to infinity in the spacetime $\mathcal{M}_{2}$.
\item \emph{Case 3}~~$M_{1}<M_{2}$: The critical impact parameter satisfies $b_{\rm c_{1}} < b_{\rm c_{2}}$. The light passes through the throat into the spacetime $\mathcal{M}_{2}$ , and then returns to the spacetime $\mathcal{M}_{1}$. Note that some additional lights from $\mathcal{M}_{2}$ will enter $\mathcal{M}_{1}$ and reach infinity in this scenario.
\end{itemize}
The impact parameters of two spacetimes satisfy \cite{65}
\begin{equation}
\label{2-10}
\frac{b_{1}}{b_{2}} = \sqrt{\frac{f_{2}(R)}{f_{1}(R)}} \equiv H.
\end{equation}
To summarize, when the impact parameter satisfies $H b_{\rm c_{\rm 2}} < b_{1} < b_{\rm c_{\rm 1}}$, the light enters the spacetime $\mathcal{M}{1}$ from infinity, traverses through the throat into the spacetime $\mathcal{M}_{2}$, returns back to $\mathcal{M}_{1}$, and eventually exits to infinity. In our subsequent discussion, we will primarily focus on \emph{Case 3}, as it elucidates the movement of photons that pass through the throat.

\par
\begin{figure*}[tbp]
\centering
  \includegraphics[width=4cm,height=4cm]{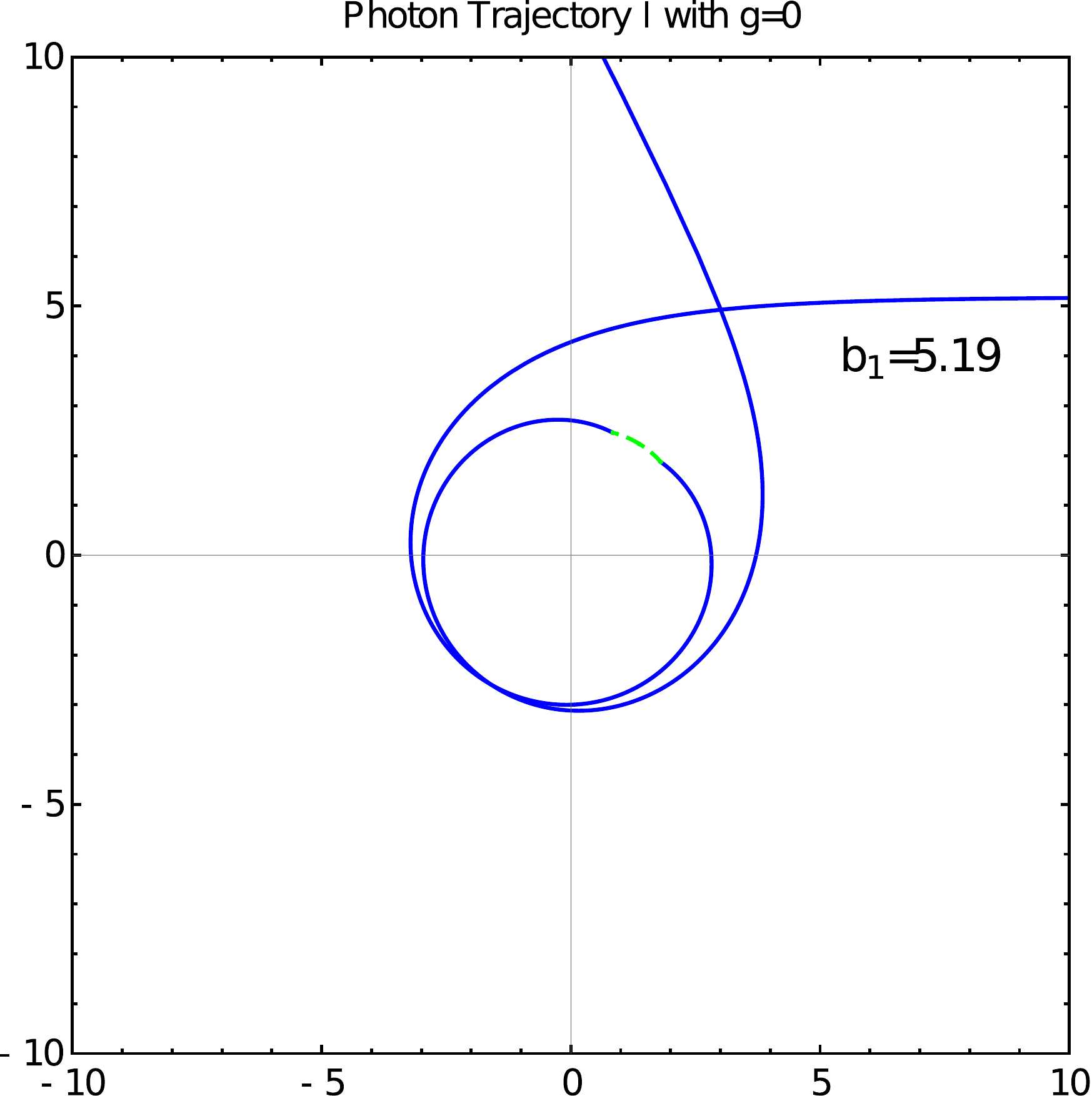}
  \hspace{0.5cm}
  \includegraphics[width=4cm,height=4cm]{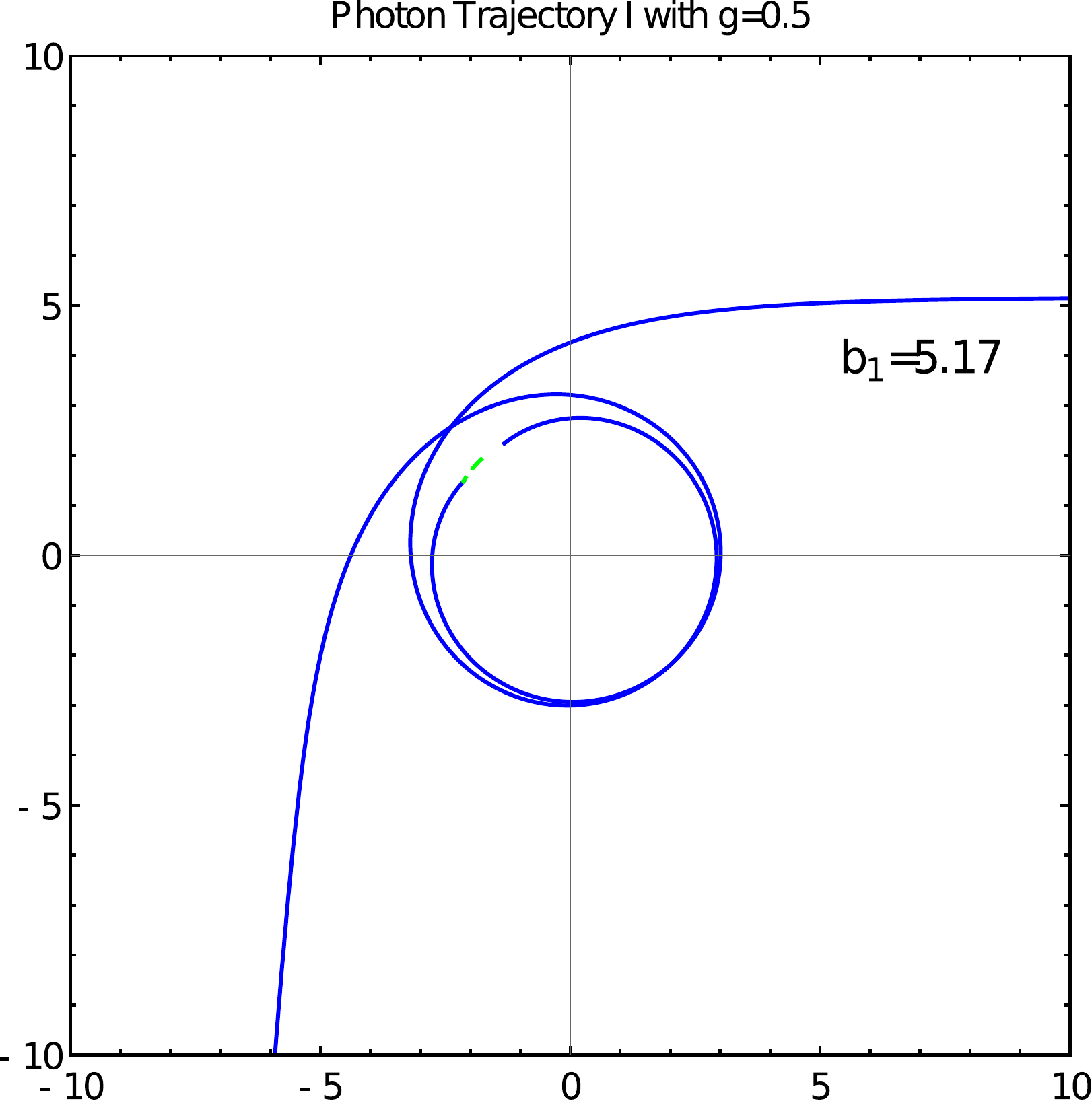}
  \hspace{0.5cm}
  \includegraphics[width=4cm,height=4cm]{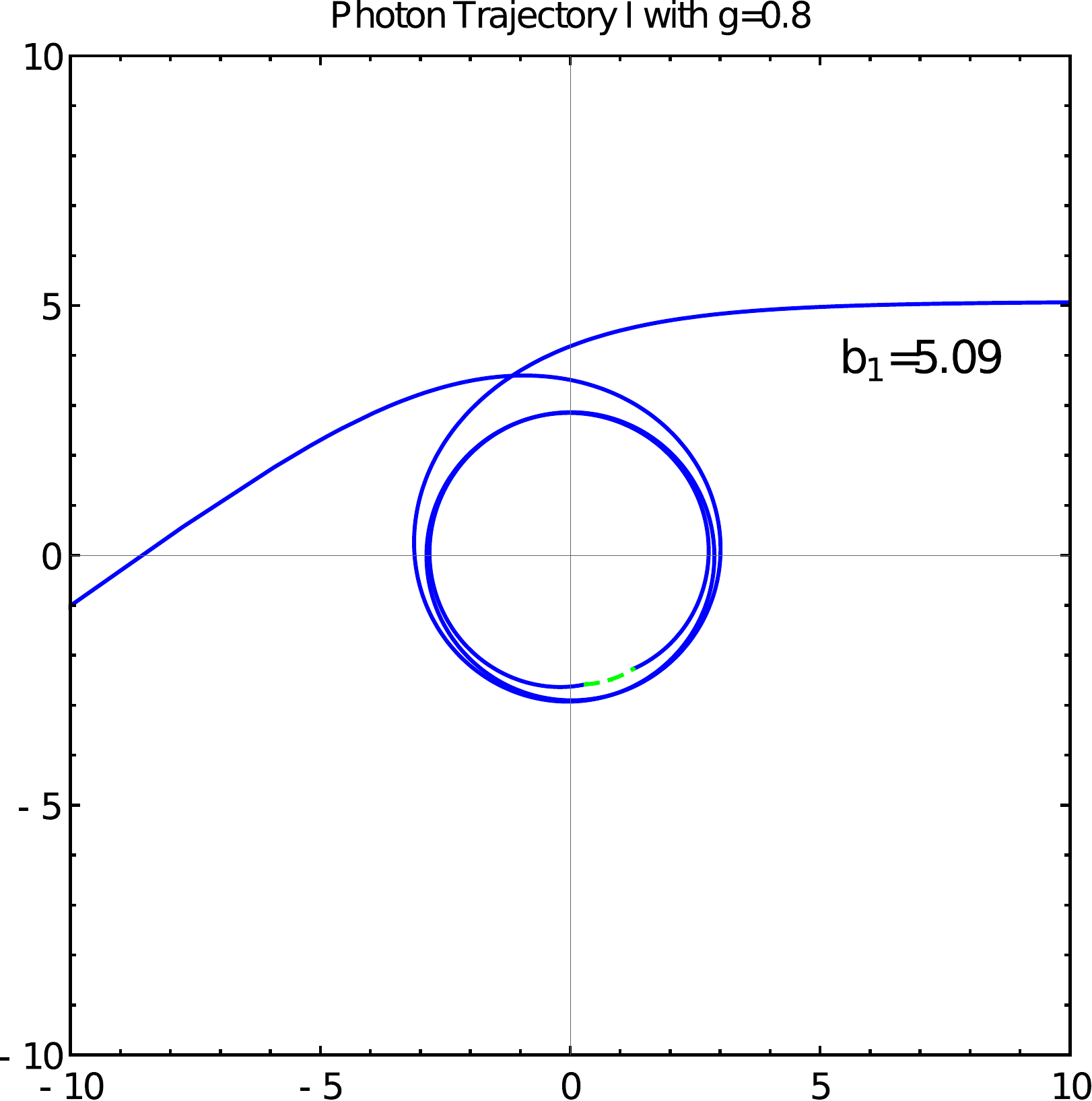}
  \hspace{0.5cm}
  \includegraphics[width=4cm,height=4cm]{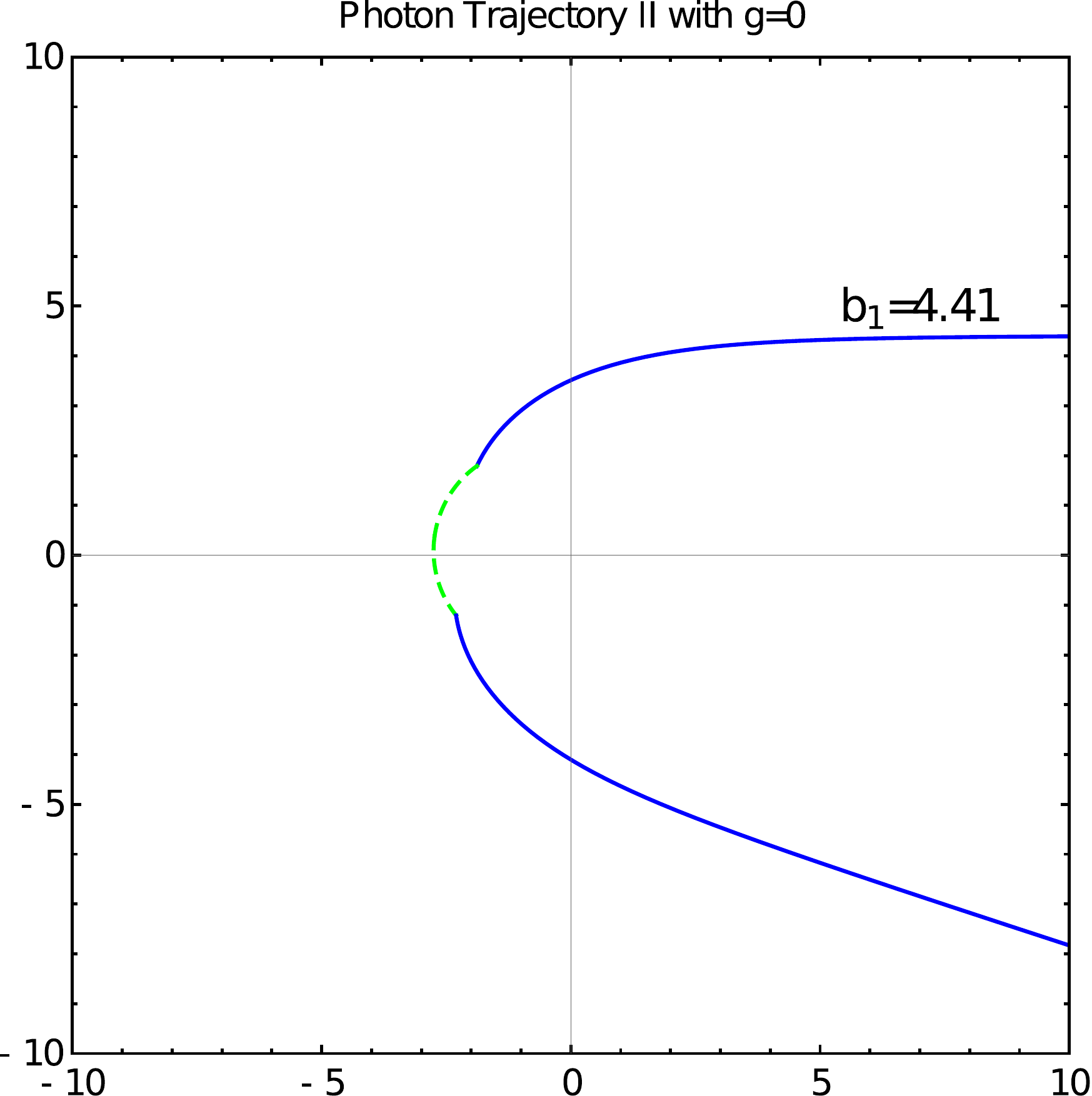}
  \hspace{0.5cm}
  \includegraphics[width=4cm,height=4cm]{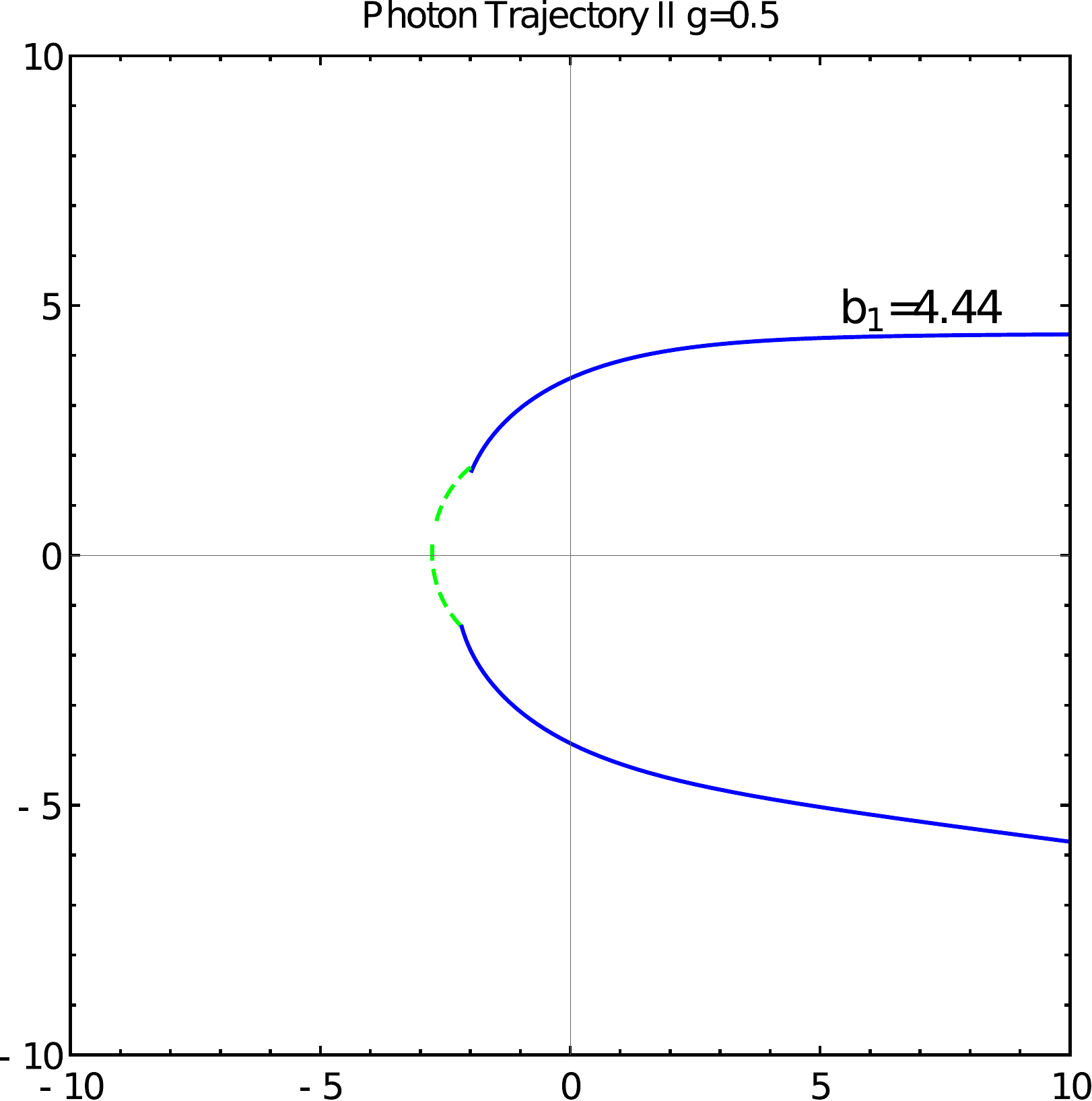}
  \hspace{0.5cm}
  \includegraphics[width=4cm,height=4cm]{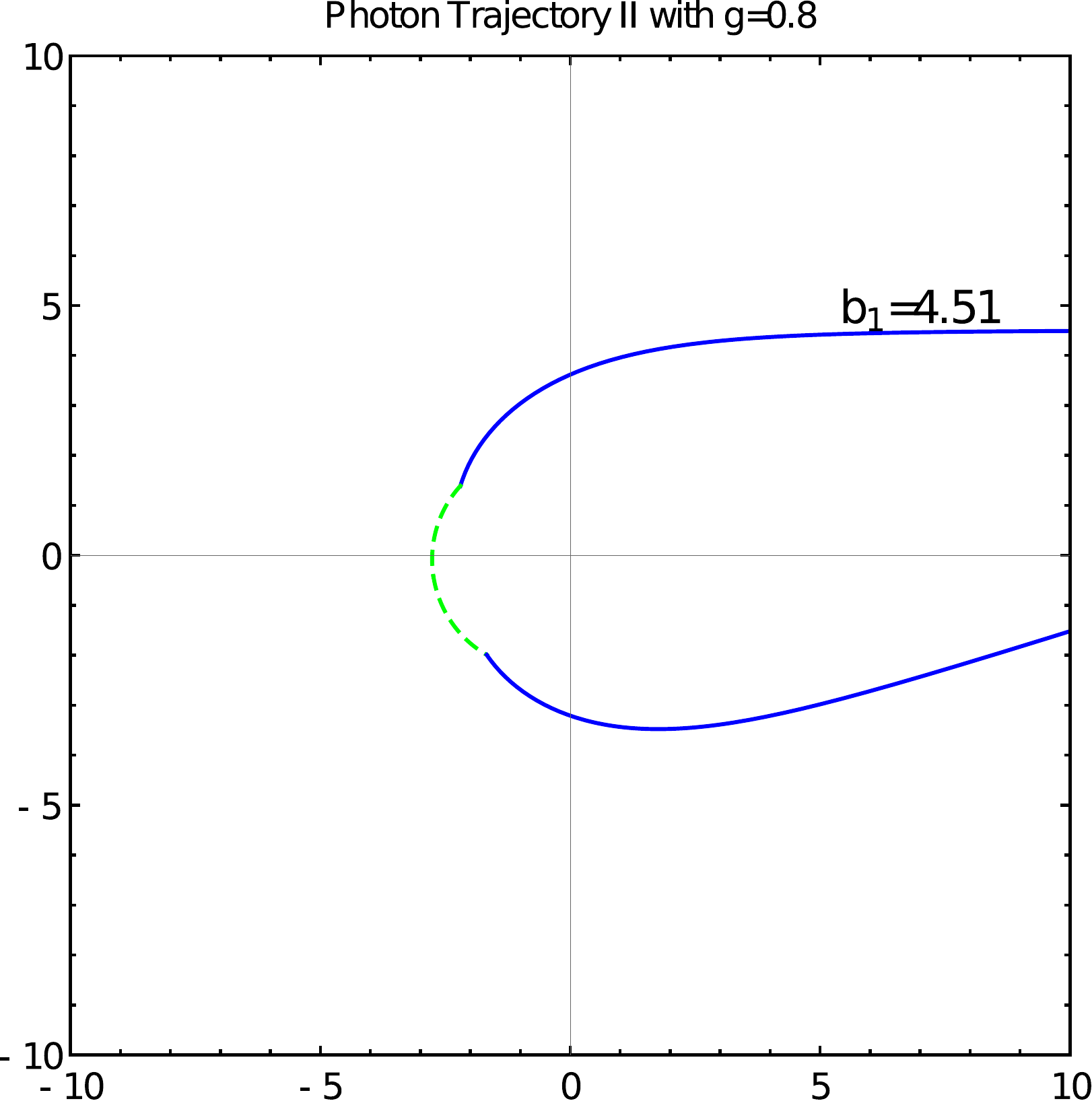}
  \hspace{0.5cm}
  \includegraphics[width=4cm,height=4cm]{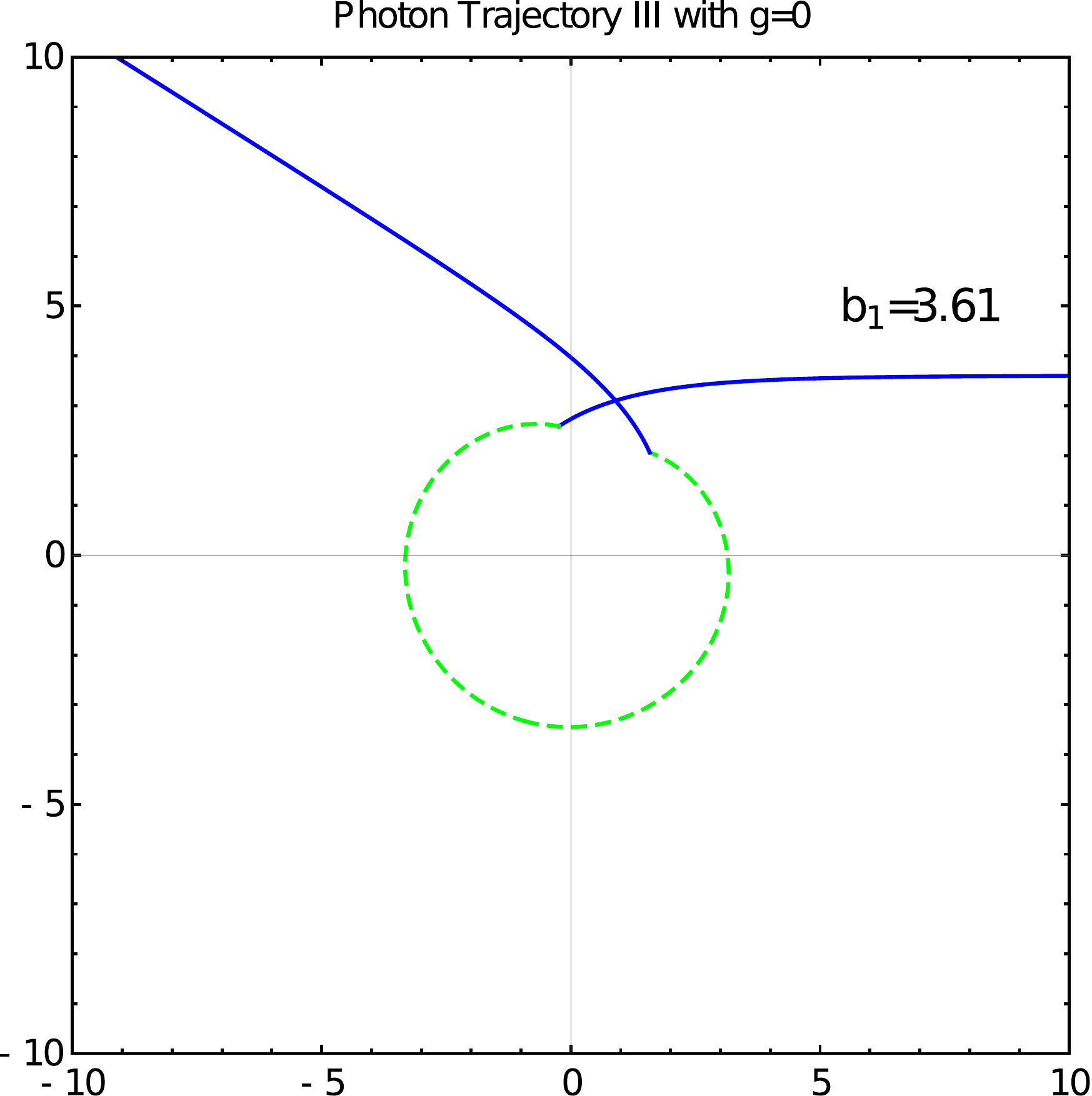}
  \hspace{0.5cm}
  \includegraphics[width=4cm,height=4cm]{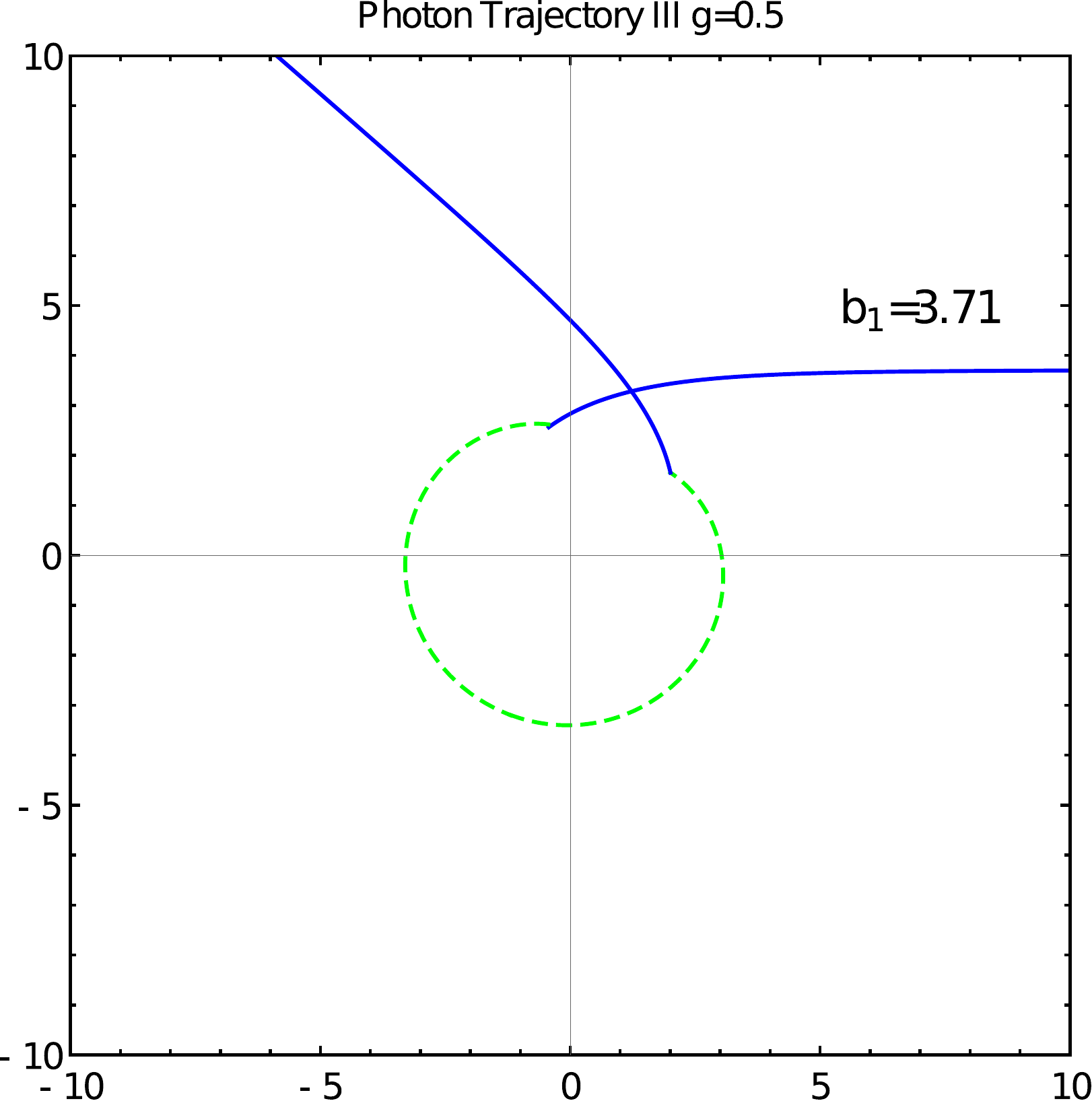}
  \hspace{0.5cm}
  \includegraphics[width=4cm,height=4cm]{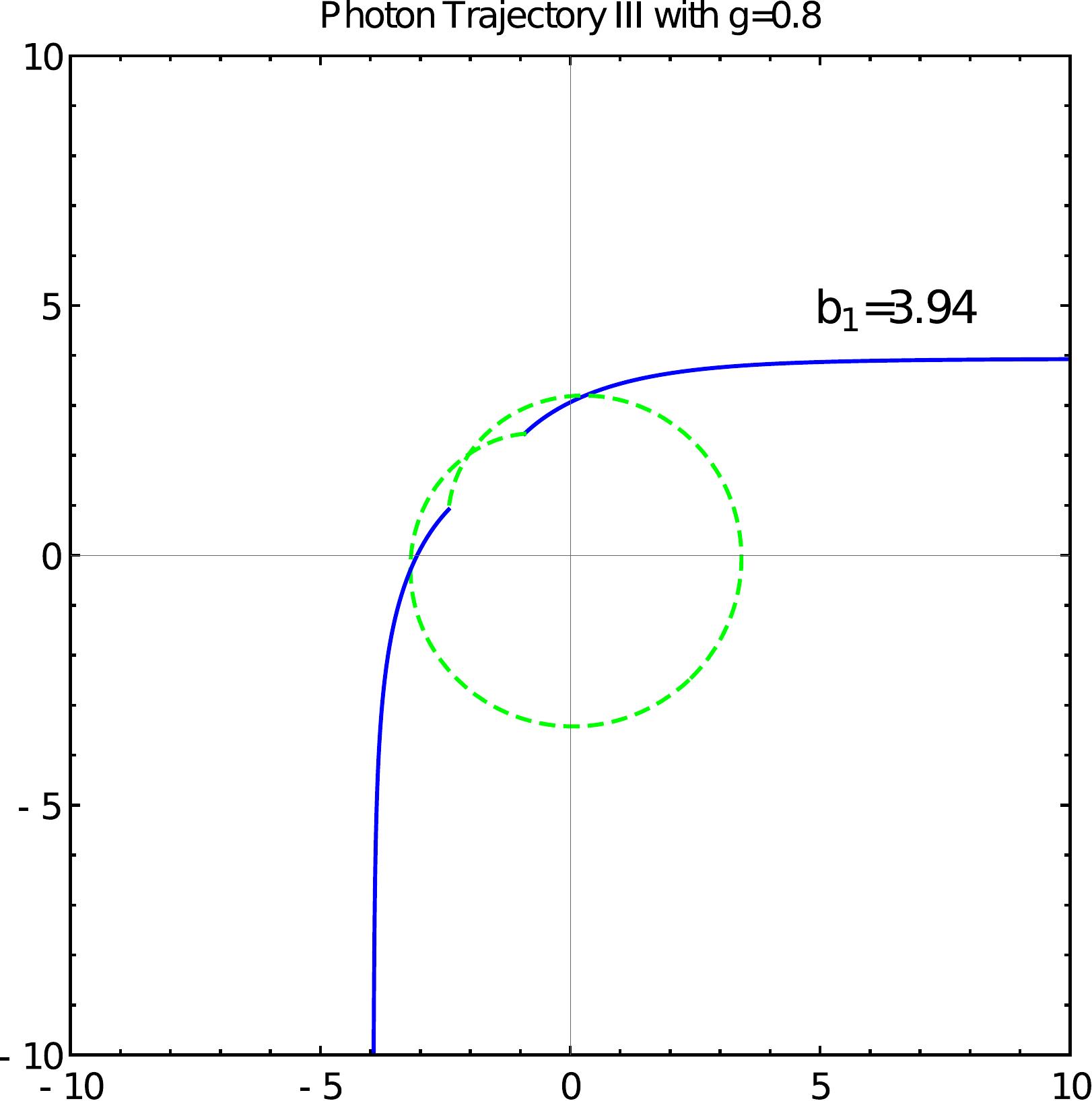}
\caption{\label{fig:3} Light trajectories in the polar coordinates $(r,\phi)$ with the impact parameters in the range $H b_{\rm c_{2}} < b_{1} < b_{\rm c_{\rm 1}}$. The blue curves represent the light trajectories in the spacetime $\mathcal{M}_{1}$ ($M_{1}=1$), and the green dashed curves represent the light trajectories in the spacetime $\mathcal{M}_{2}$ ($M_{2}=1.2$). {\em Left Panel}-- magnetic charge $g=0$ (Schwarzschild TSW scenario), {\em Middle Panel}-- magnetic charge $g=0.5$ and {\em Right Panel}-- magnetic charge $g=0.8$. We take throat radius $R=2.6$.}
\end{figure*}

By utilizing a ray-tracing code, we can effectively determine the trajectory of light within a TSW with a Hayward profile. The light trajectory can be described by the orbit equation, denoted as Eq. (\ref{2-6}), and can be reexpressed as follows:
\begin{equation}
\label{3-1}
\frac{1}{b_{\rm i}^{2}}-\frac{f_{\rm i}(r_{\rm i})}{r_{\rm i}^{2}} = \frac{1}{r_{\rm i}^{4}}\Big(\frac{{\rm d} r_{\rm i}}{{\rm d} \phi_{\rm i}}\Big)^{2}.
\end{equation}
By introducing a parameter $u \equiv 1/r$, one can get
\begin{equation}
\label{3-2}
\Omega_{\rm i}(u_{\rm i}) \equiv \frac{{\rm d} u_{\rm i}}{{\rm d} \phi_{\rm i}} = \sqrt{\frac{1}{b_{\rm i}^{2}}-u_{\rm i}^{2}\Bigg[1-\frac{2M_{\rm i}}{u_{\rm i}^{2}(g^{3}+\frac{1}{u_{\rm i}^{3}})}\Bigg]}.
\end{equation}
To provide an illustration, we present in Fig. \ref{fig:3} the light trajectories within a TSW with a Hayward (Schwarzschild) profile, considering different values of the magnetic charge. Assuming that the light originates from the far right within spacetime $\mathcal{M}_{1}$, we observe that smaller impact parameters result in more extensive regions of light motion within spacetime $\mathcal{M}_{2}$. This implies that light rays with smaller impact parameters experience a greater influence from the gravitational field and exhibit more significant deflection as they traverse through the TSW.

\par
For impact parameters $b_{1}>b_{\rm c_{1}}$, the inflection point of the light trajectory within spacetime $\mathcal{M}_{1}$ corresponds to the minimally positive real root of $\Omega_{1}(u_{1})$, which we denote as $u_{1}^{\rm \min}$. Utilizing Eq. (\ref{3-2}), the total change of azimuthal angle within spacetime $\mathcal{M}_{1}$ can be expressed as follows:
\begin{equation}
\label{3-3}
\phi_{1}(b_{1}) = 2 \int_{0}^{u_{1}^{\rm \min}} \frac{{\rm d}u_{1}}{\sqrt{\Omega_{1}(u_{1})}},~~~~~~~b_{1}>b_{\rm c_{1}}.
\end{equation}
This situation is similar to that of the BH. When $H b_{\rm c_{2}} < b_{1} < b_{\rm c_{\rm 1}}$, the change in azimuthal angle (with the throat outside) in the spacetime $\mathcal{M}_{1}$ can be given as
\begin{equation}
\label{3-4}
\phi_{1}^{*}(b_{1}) = \int_{0}^{1/R} \frac{{\rm d}u_{1}}{\sqrt{\Omega_{1}(u_{1})}},~~~~~~~~b_{1}<b_{\rm c_{1}}.
\end{equation}
Subsequently, the light ray enters the spacetime $\mathcal{M}_{2}$ through the throat. In this region, the inflection point of the light trajectory within spacetime $\mathcal{M}_{2}$ corresponds to the largest positive real root of $\Omega_{2}(u_{2})$, denoted as $u_{2}^{\rm \max}$. By utilizing Eq. (\ref{3-2}), we can calculate the change in azimuthal angle within spacetime $\mathcal{M}_{2}$, which is given by
\begin{equation}
\label{3-5}
\phi_{2}(b_{2}) = 2 \int_{u_{2}^{\rm \max}}^{1/R} \frac{{\rm d} u_{2}}{\sqrt{\Omega_{2}(u_{2})}},~~~~~~~b_{2}>b_{\rm c_{2}}.
\end{equation}

\section{\textbf{Optical appearance of a TSW with a Hayward profile}}
\label{sec:4}
\subsection{\textbf{Light rays classification}}
\label{sec:4-1}
Gralla $et~al.$ have proposed that radiation originates from an optically and geometrically thin accretion disk situated in the equatorial plane of the BH \cite{27}. Based on the number of times the light intersects the accretion disk and the number of light orbits ($n \equiv \phi / 2 \pi$), the light rays can be classified into the following categories:
\begin{itemize}
\item \emph{Case 1}~~When $n > 1/4$, the light falls onto the front side of the disk and intersects the equatorial plane only once.
\item \emph{Case 2}~~When $n > 3/4$, the light penetrates through the thin disk, passing to the back side, and intersects the equatorial plane twice.
\item \emph{Case 3}~~When $n > 5/4$, the light reaches the front side of the accretion disk once again, intersecting the equatorial plane more than three times. These additional intersections contribute to an enhanced brightness observed by an observer.
\end{itemize}

\par
In the case of a TSW with a Hayward profile, we consider the presence of a static observer and a thin accretion disk situated in the spacetime $\mathcal{M}_{1}$. The static observer is positioned at the north pole, while the accretion disk resides in the equatorial plane and emits radiation isotropically in the rest frame of the static worldlines. When the light enters spacetime $\mathcal{M}_{1}$ and subsequently falls into spacetime $\mathcal{M}_{2}$ through the throat, the number of orbits within the TSW can be defined as follows \cite{66}:
\begin{eqnarray}
\label{4-1}
&&n_{1}(b_{1}) = \frac{\phi_{1}(b_{1})}{2\pi},\\
&&n_{2}(b_{2}) = \frac{\phi_{1}^{*}(b_{1}) + \phi_{2}(b_{1}/H)}{2\pi},\\
&&n_{3}(b_{1}) = \frac{2 \phi_{1}^{*}(b_{1}) + \phi_{2}(b_{1}/H)}{2\pi}.
\end{eqnarray}
In the case where the number of orbits is $n_{1}$, it corresponds to the scenario similar to that of a BH. If $n_{2}<3/4$ and $n_{3}>3/4$, the outgoing reflected light within spacetime $\mathcal{M}_{1}$ falls onto the back side of the accretion disk. Similarly, if $n{2}<5/4$ and $n_{3}>5/4$, the outgoing reflected light within spacetime $\mathcal{M}_{1}$ falls onto the front side of the disk.

\par
\begin{figure*}[tbp]
\centering
  \includegraphics[width=7.5cm,height=5cm]{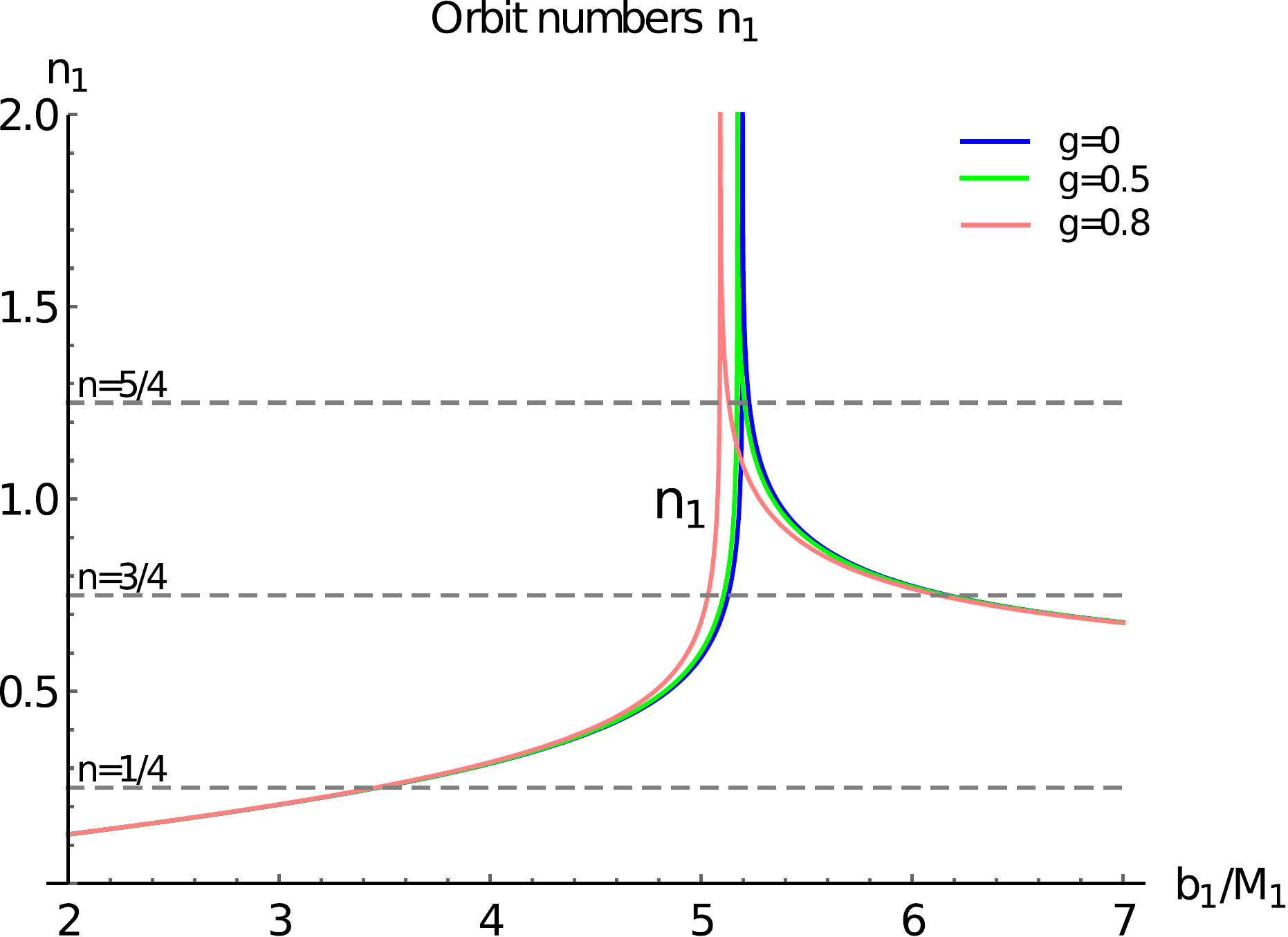}
  \includegraphics[width=7.5cm,height=5cm]{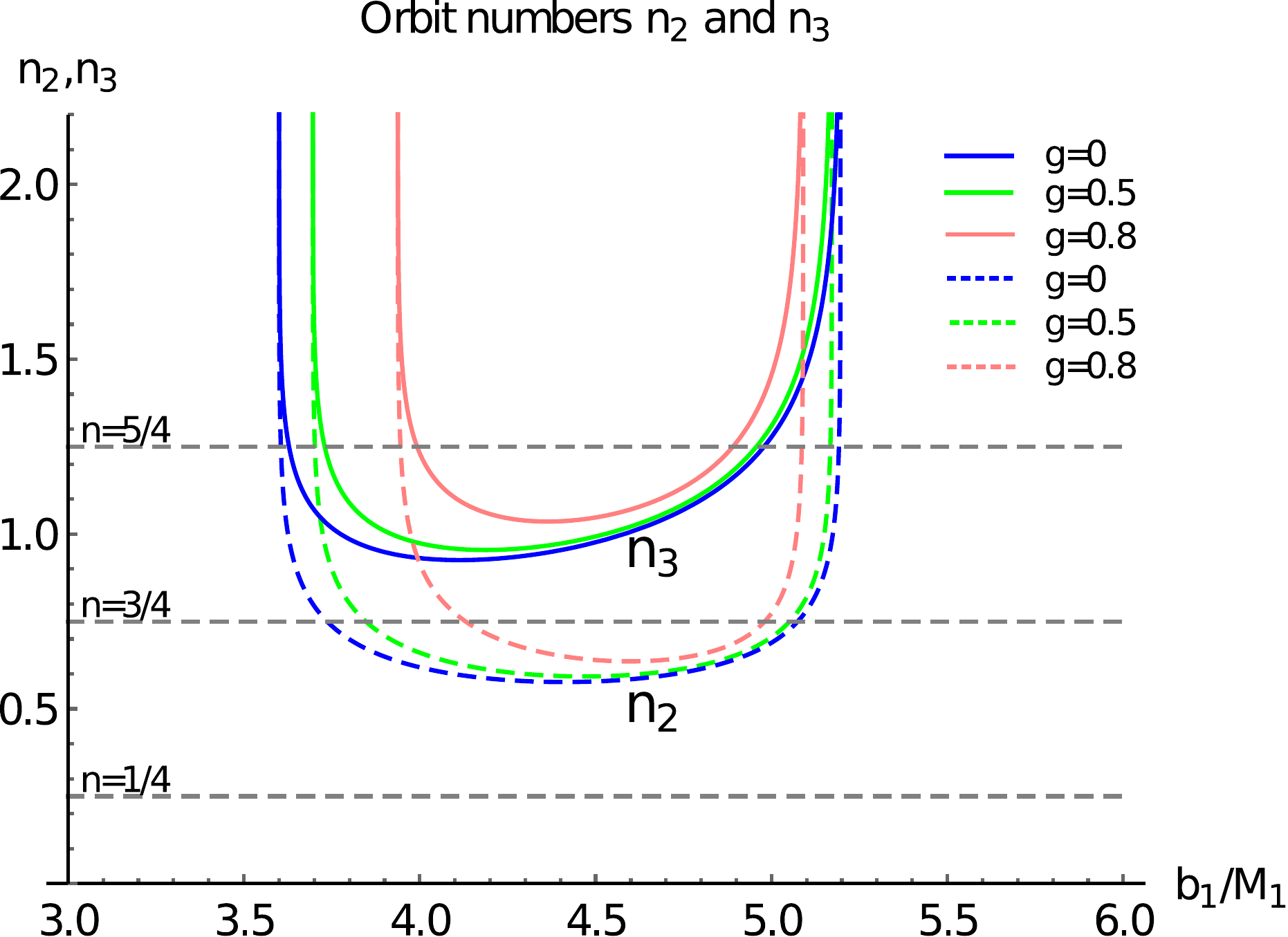}
\caption{\label{fig:4} Total orbit number as a function of the impact parameter for a TSW with a Hayward profile. The BHs mass are taken as $M_{1}=1$, $M_{2}=1.2$ and the throat radius is taken as $R=2.6$. The blue, green, and pink curves represent the magnetic charges $g=0$ (Schwarzschild TSW scenario), $g=0.5$ and $g=0.8$, respectively.}
\end{figure*}

Figure \ref{fig:4} illustrates the total number of orbits as a function of the impact parameter for a TSW with a Hayward profile. It can be observed that the orbit function $n_{1}$ of the TSW with a Hayward profile closely resembles that of a Hayward BH, implying that the image seen by an observer of the TSW contains the image of the BH. However, it should be noted that additional orbit functions $n_{2}$ and $n_{3}$ appear in the TSW scenario, indicating the presence of additional ring structures in the TSW image. Moreover, it has been observed that an increase in the value of the parameter $g$ leads to a reduction in the range of the impact parameter $b$ compared to a TSW with a Schwarzschild profile ($g=0$). This suggests that the additional ring structures are drawn inward towards the BH as the magnetic charge increases.

\subsection{\textbf{Observed intensity and transfer functions}}
\label{sec:4-2}
\par
As mentioned earlier, the light can gain additional brightness through multiple intersections with the accretion disk. The total observed intensity ($\rm ergs^{-1}cm^{-2}str^{-1}Hz^{-1}$) is the sum of these intensities. According to Liouville's theorem, $I_{\rm e}/({\upsilon_{\rm e}})^{3}$ is conserved along the direction of light propagation, where $I_{\rm e}$ is the specific intensity of the radiation and $\upsilon_{\rm e}$ is the frequency of the radiation in a static frame. An observer located at infinity receives a specific intensity $I_{\rm o}$ with a redshifted frequency $\upsilon_{\rm o} \equiv \sqrt{f} \upsilon_{\rm e}$. Taking into account the conservation of $I/\upsilon^{3}$ along a ray, we obtain the following relation:
\begin{equation}
\label{4-2-1}
\frac{I_{\rm o}}{\upsilon_{\rm o}^{3}}=\frac{I_{\rm e}}{\upsilon_{\rm e}^{3}}.
\end{equation}
For a single frequency, the observed specific intensity is
\begin{eqnarray}
\label{4-2-2}
I_{\rm o}(r)= f(r)^{{3}/{2}}I_{\rm e}(r).
\end{eqnarray}
By integrating over the whole range of received frequencies, the total observed intensity can be written as
\begin{eqnarray}
\label{4-2-3}
I_{\rm S}(r)&&=\int I_{\rm o}(r) {\rm d} \upsilon_{\rm o}\nonumber \\
&&= \int f(r)^{2} I_{\rm e}(r) {\rm d} \upsilon_{\rm e}= f(r)^{2} I_{\rm R}(r),
\end{eqnarray}
where $I_{\rm R}(r) \equiv \int I_{\rm e}(r) {\rm d} \upsilon_{\rm e}$ is the total radiation intensity from the thin accretion disk. The total observed intensity is
\begin{eqnarray}
\label{4-2-4}
I_{\rm obs}(b)= \sum \limits_{n} f(r)^{2} I_{\rm R}|_{r=r_{\rm n}(b)},
\end{eqnarray}
in which $r_{\rm n}(b)$ represents the transfer function, which provides the radial coordinate of the $n_{\rm th}$ intersection between the light ray with impact parameter $b$ and the accretion disk. The slope of the transfer function, denoted as ${\rm d}r/{\rm d}b$, is defined as the (de)magnification factor \cite{27}. In the case of a BH, the first transfer function ($n=1$) corresponds to the ``direct emission'' from the accretion disk, while the second ($n=2$) and third ($n=3$) transfer functions represent the ``lensing ring'' and ``photon ring'', respectively.

\par
\begin{figure*}[tbp]
\centering
  \includegraphics[width=5cm,height=4.5cm]{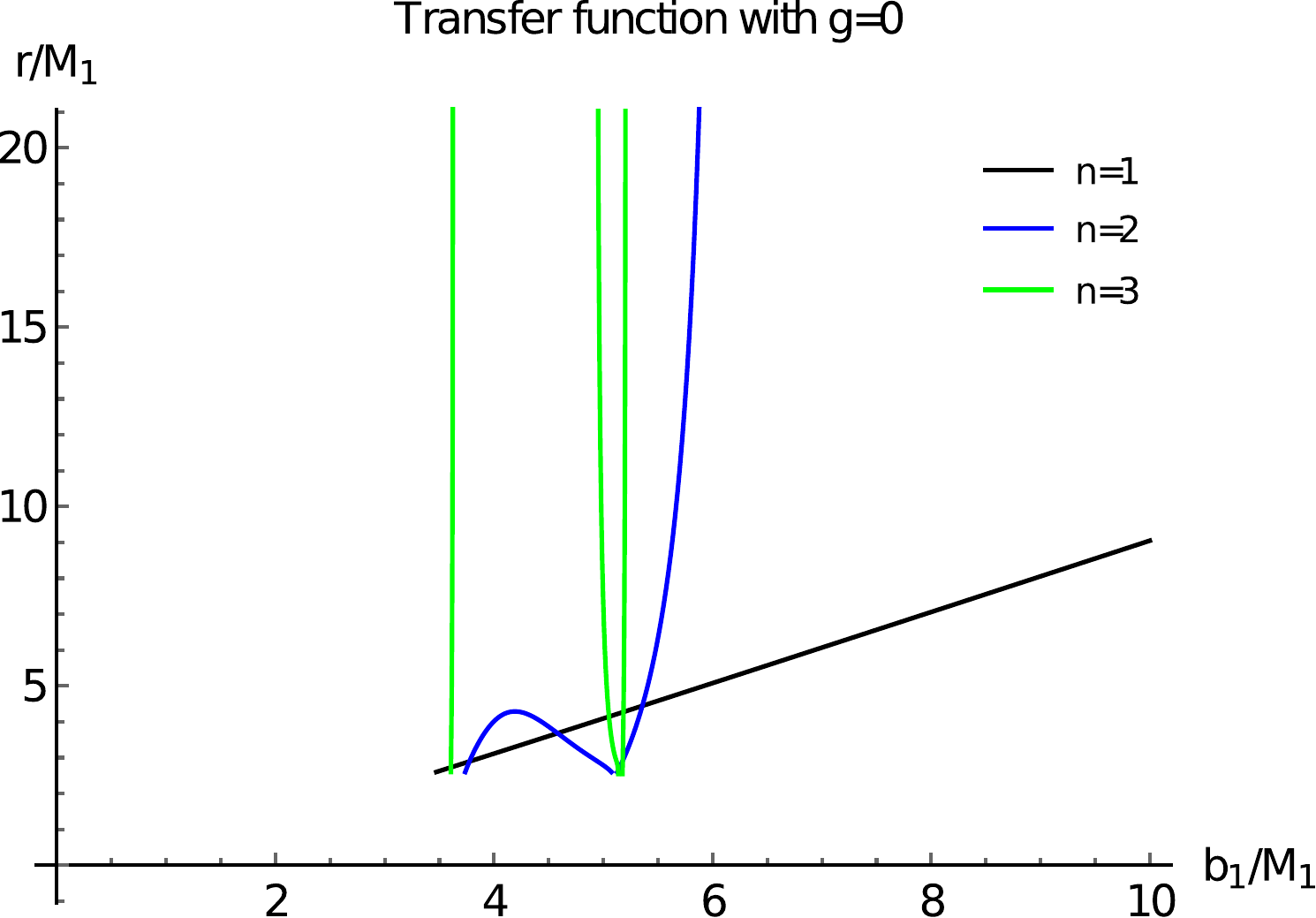}
  \includegraphics[width=5cm,height=4.5cm]{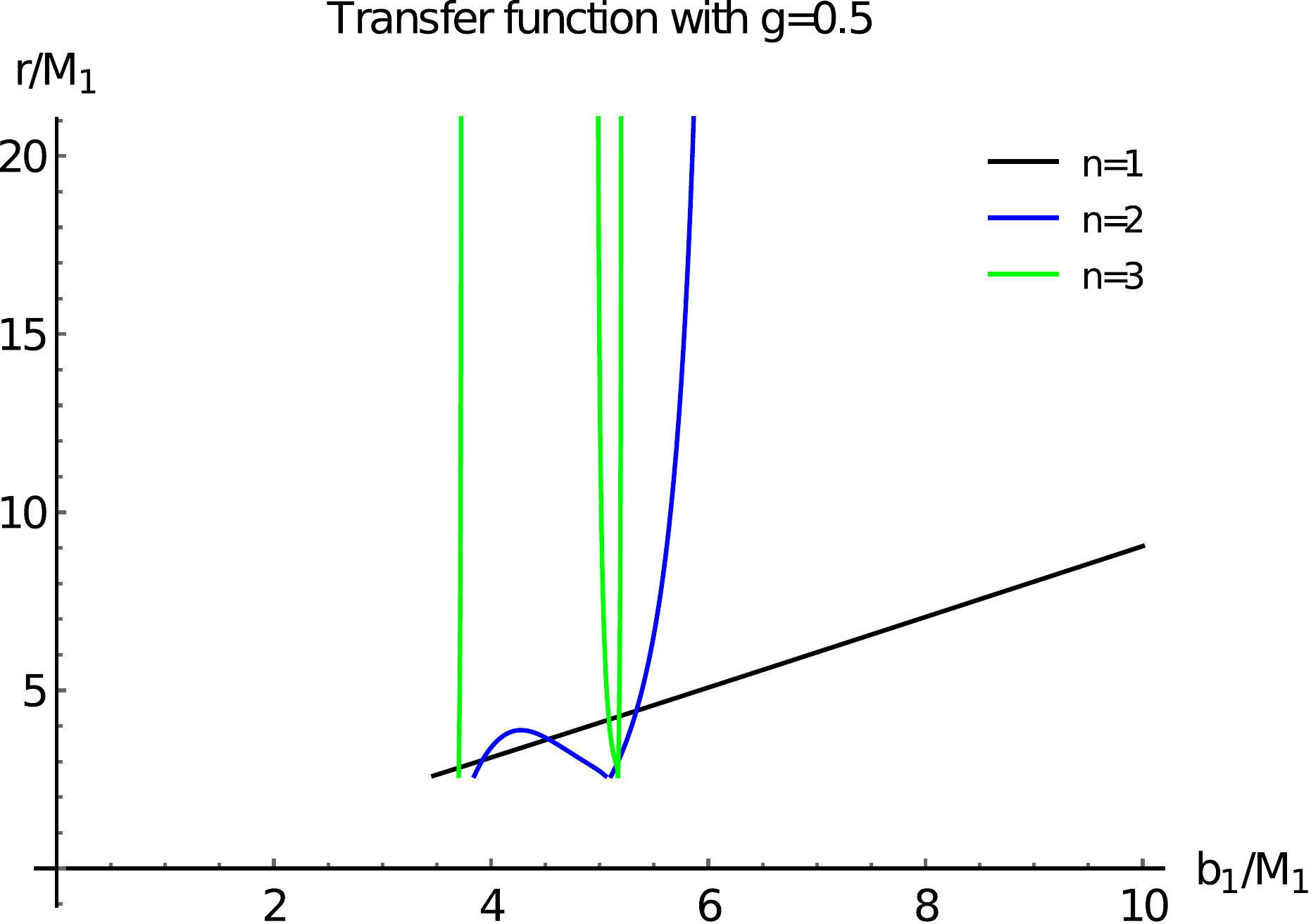}
  \includegraphics[width=5cm,height=4.5cm]{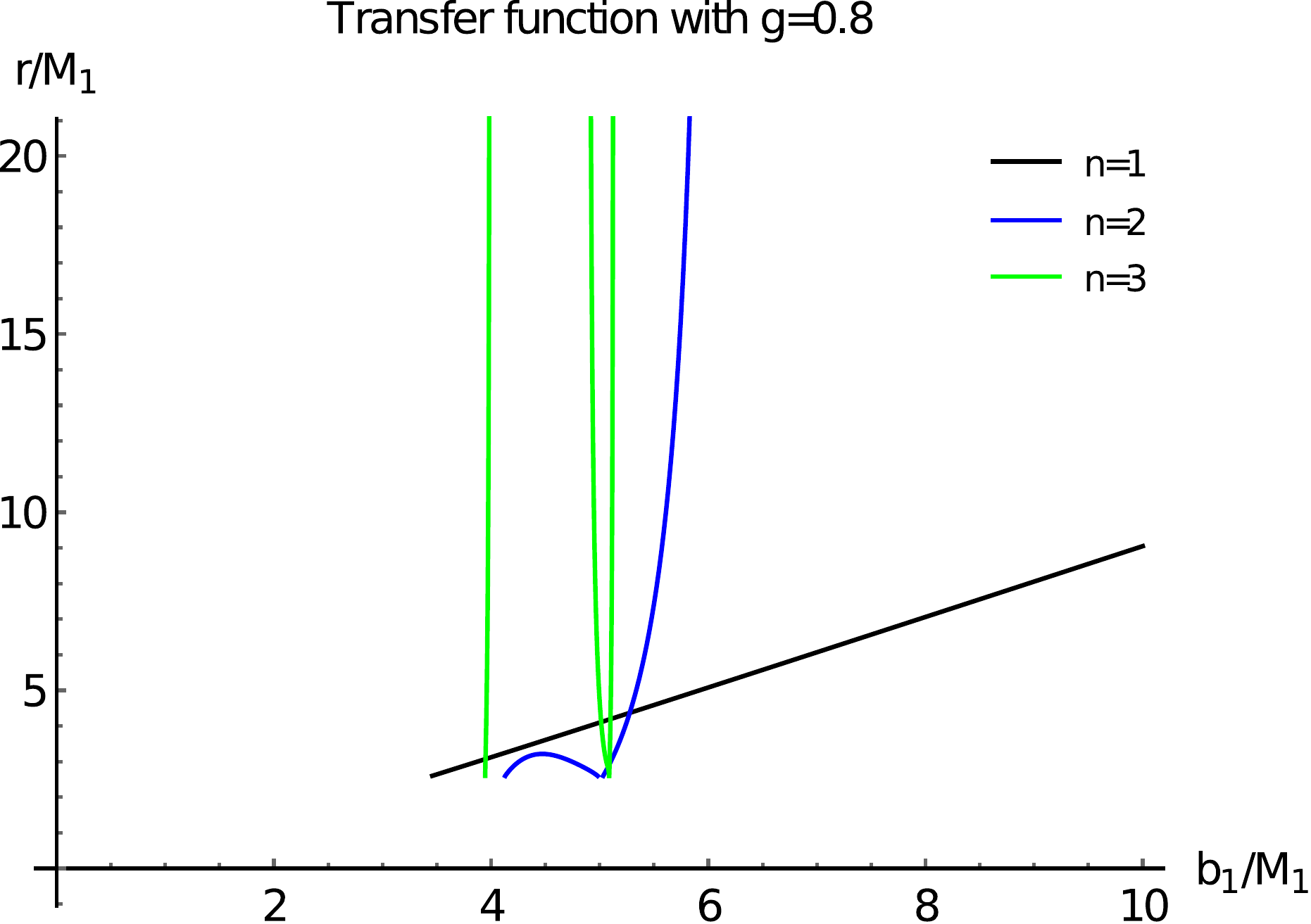}
\caption{\label{fig:5} Transfer functions of a TSW with a Hayward profile under the magnetic charges of $g=0$ (Schwarzschild TSW scenario) [{\em Panel (a)}], $g=0.5$ [{\em Panel (b)}], and $g=0.8$ [{\em Panel (c)}]. The black, blue, and green curves are for the direct emission, lensing band, and photon ring group, respectively. The BHs mass are taken as $M_{1}=1$, $M_{2}=1.2$, and the throat radius is taken as $R=2.6$.}
\end{figure*}

Figure \ref{fig:5} showcases the transfer function $r_{\rm n}(b)$ as a function of the impact parameter $b$ for several representative magnetic charges. The black lines represent the first transfer function ($n=1$), which closely resembles that of the Hayward BH. It can be observed that $r_{1}(b)$ is proportional to $b$ and exhibits a slope of approximately 1. This indicates that the direct emission contributes the most to the total observed flux. The blue lines correspond to the second transfer function ($n=2$). It can be seen that the curve has an irregular shape with a slightly wider range of impact parameters, connecting at the lower end of the asymptotic curve. This illustrates that the observer perceives a ``lensing band'' in the case of a TSW. The green lines represent the third transfer function ($n=3$). In addition to the usual third transfer function (which appears as an almost vertical line near $b_{\rm c_{1}}$), two new third transfer functions are present. One of them is located close to $H b_{\rm c_{2}}$ and exhibits the same slope as the usual third transfer function. The other one is situated to the left of $b_{\rm c_{1}}$, and its slope is smaller than the usual third transfer function but greater than the usual second transfer function. Consequently, the observer perceives a ``photon ring group'' in the image.

\subsection{\textbf{Optical appearance of a TSW with a Hayward profile}}
\label{sec:4-3}
\par
In order to study the optical appearance of a TSW with a Hayward profile, we consider two radiation functions for the thin accretion disk. It is widely recognized that the radiation emitted by accretion disks in the Universe can be approximated by a Gaussian distribution \cite{74}. Hence, we parameterize the radiation function of the accretion disk as a Gaussian function.

\par
Given a BH mass of $1$, the radius of the innermost stable circular orbit is found to be approximately $r_{\rm isco} \simeq 6$. We will consider Model I, where the inner edge of the accretion disk is located at $r=6$, beyond which no radiation is emitted. To characterize the radiation function of the accretion disk, we adopt a Gaussian distribution function, which can be expressed as follows:
\begin{equation}
\label{4-3-1}
I_{\rm R_{1}}(r)~=~\left\{
\begin{array}{rcl}
exp\Big[{\frac{-(r-6)^{2}}{8}}\Big] ~~~~~~~~~~& & {r>6},\\
\\
0~~~~~~~~~~~~~~~~~~~~ & & {r \leq 6}.\end{array} \right.
\end{equation}

\par
Next, we consider Model II, where the radiation function falls to zero more smoothly compared to Model I, in order to provide a contrast between the two models. In this case, we assume that the inner radii at which the accretion stops radiating are given by the event horizon radius of the BH ($r_{+} \simeq 2$). The radiation function can be described as follows:
\begin{equation}
\label{4-3-2}
I_{\rm R_{2}}(r)~=~\left\{
\begin{array}{rcl}
\frac{\frac{\pi}{2}-\arctan(r-5)}{\frac{\pi}{2}-\arctan (-3)} ~~~~~~~~~~& & {r>2},\\
\\
0~~~~~~~~~~~~~~~~~~~~ & & {r \leq 2},
\end{array} \right.
\end{equation}
\begin{figure}[tbp]
\centering
  \includegraphics[width=8cm,height=5cm]{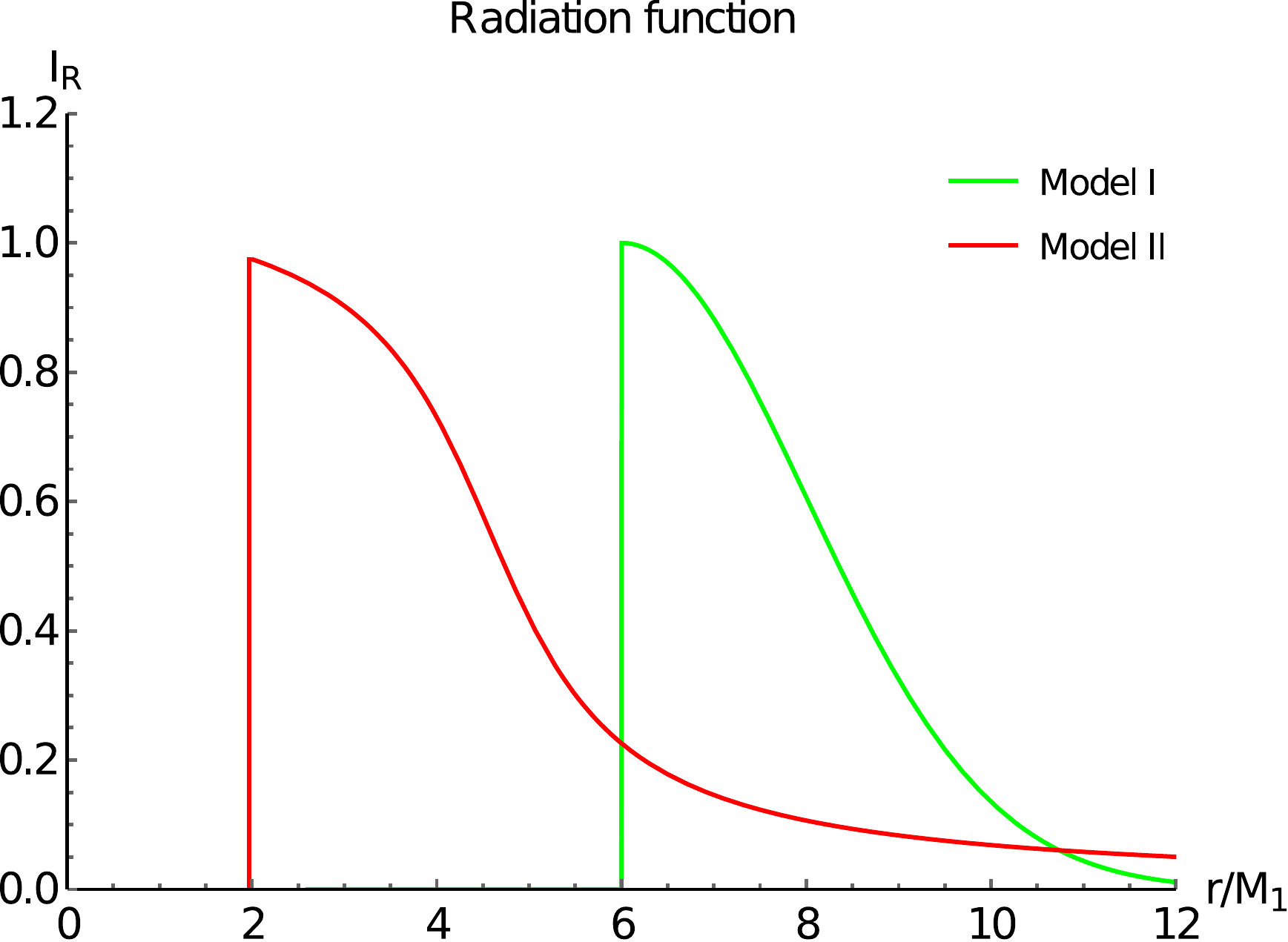}
\caption{\label{fig:6} The radiation intensity as a function of the radius. The green and red curves are for the Model I and Model II.}
\end{figure}
\begin{figure*}[tbp]
\centering
  \includegraphics[width=6cm,height=4cm]{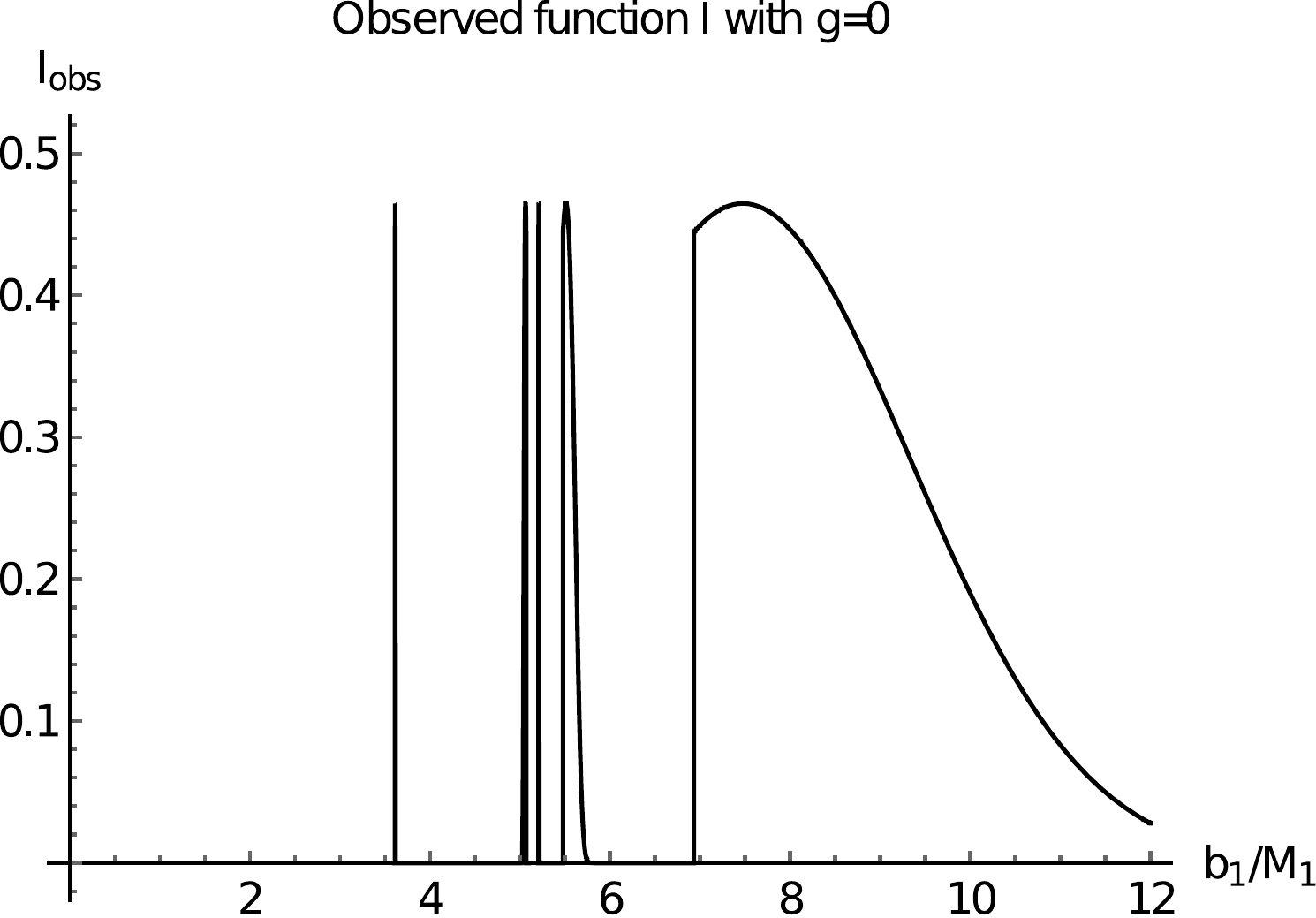}
  \hspace{0.5cm}
  \includegraphics[width=4cm,height=4cm]{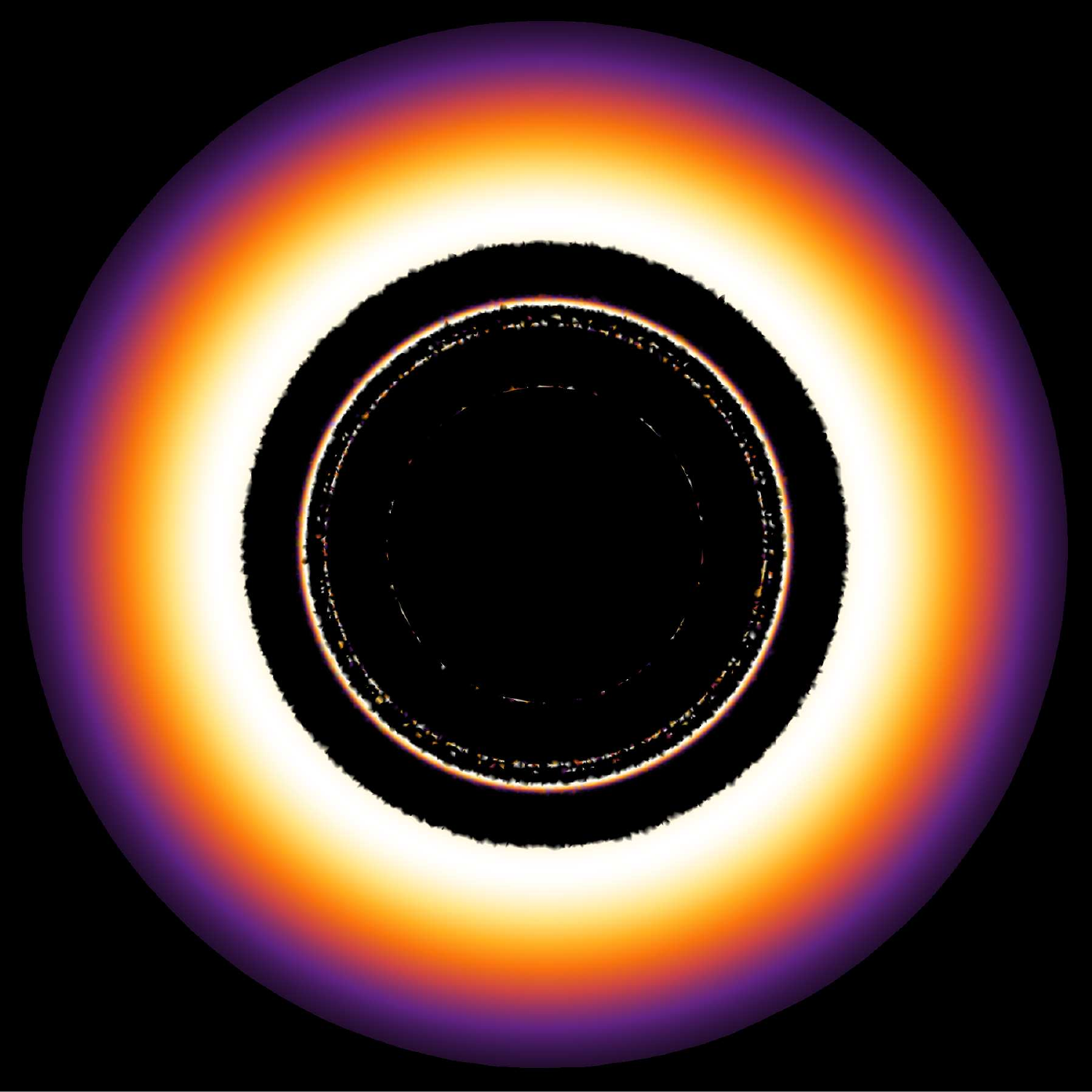}
  \hspace{0.5cm}
  \includegraphics[width=4cm,height=4cm]{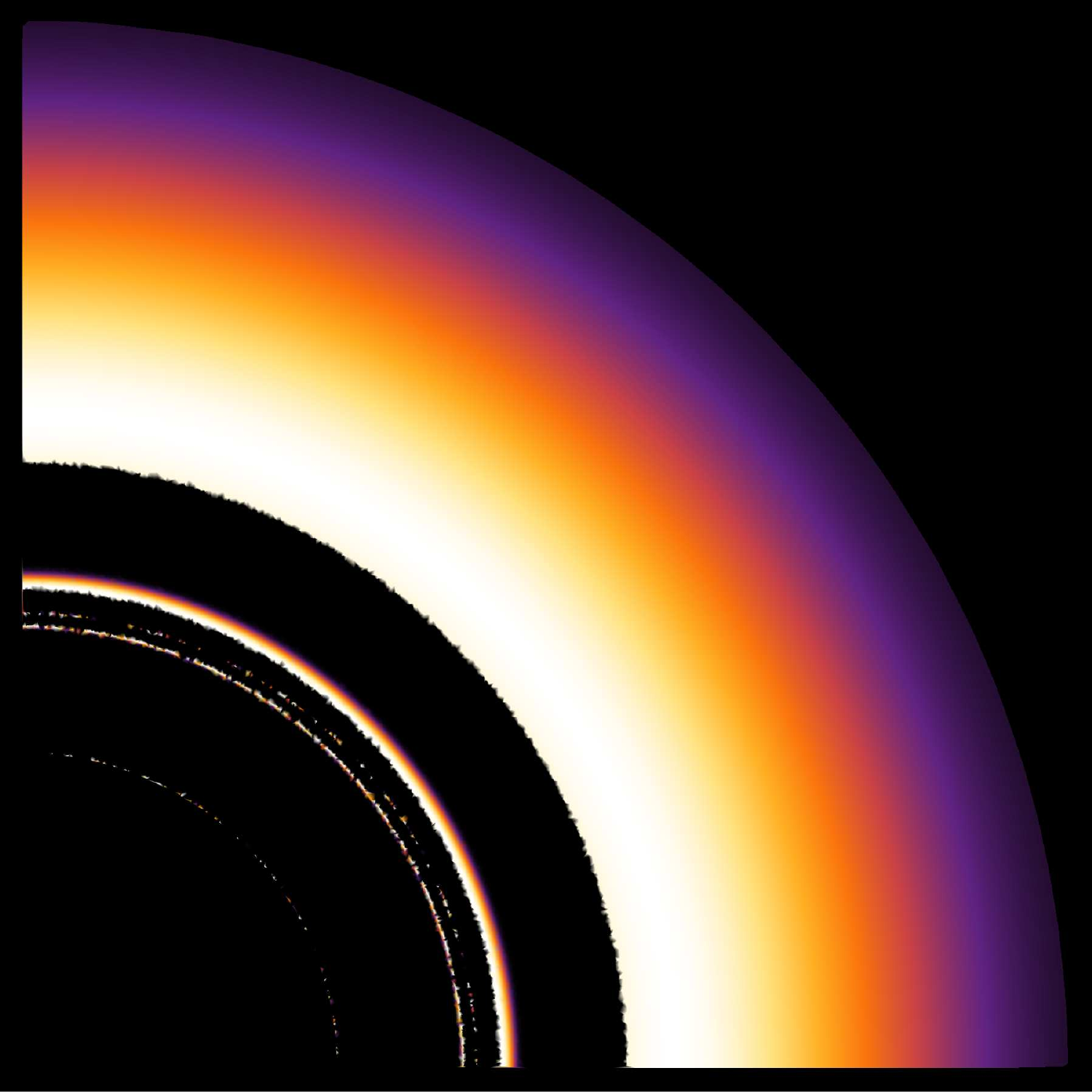}
  \hspace{0.5cm}
  \includegraphics[width=6cm,height=4cm]{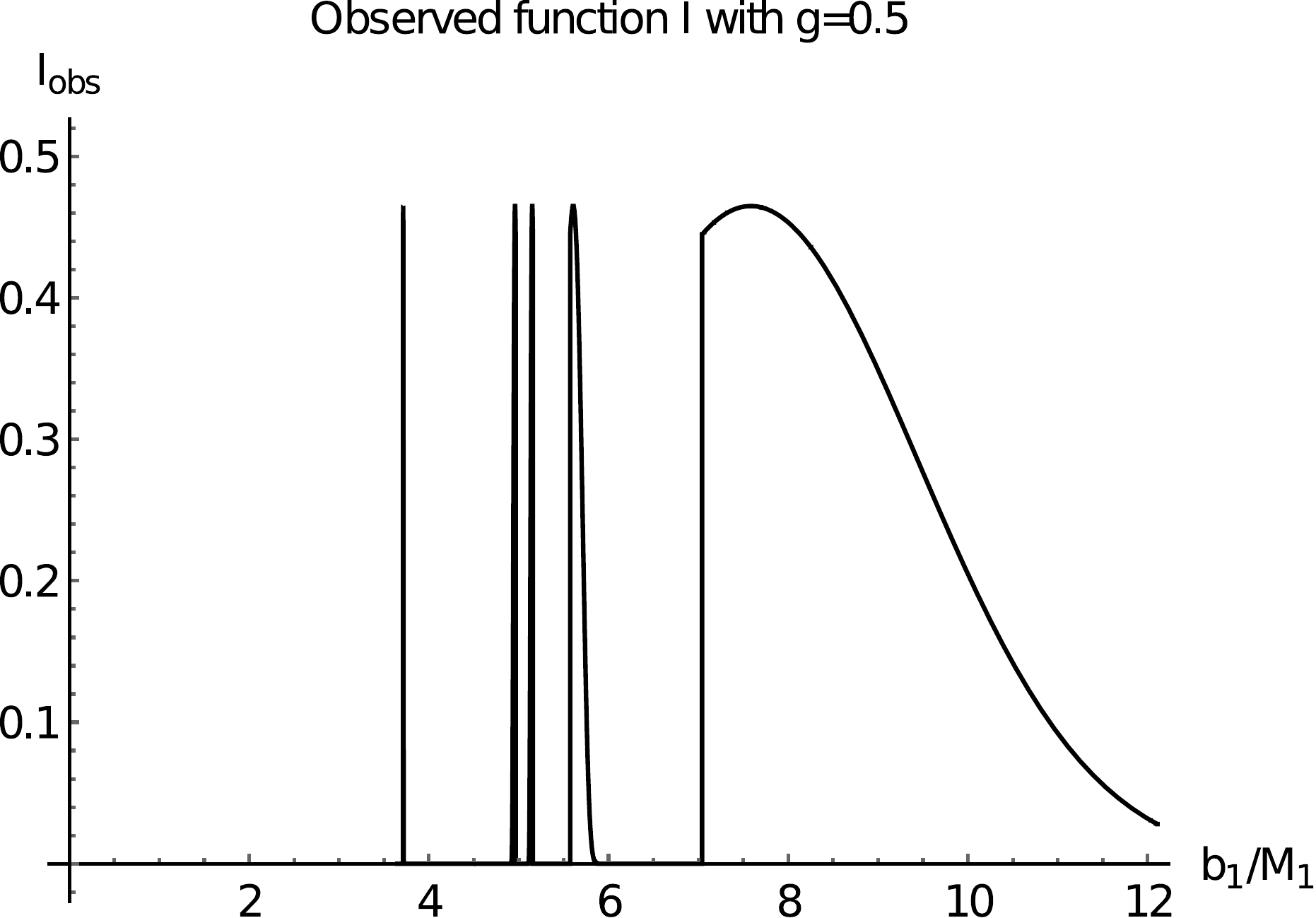}
  \hspace{0.5cm}
  \includegraphics[width=4cm,height=4cm]{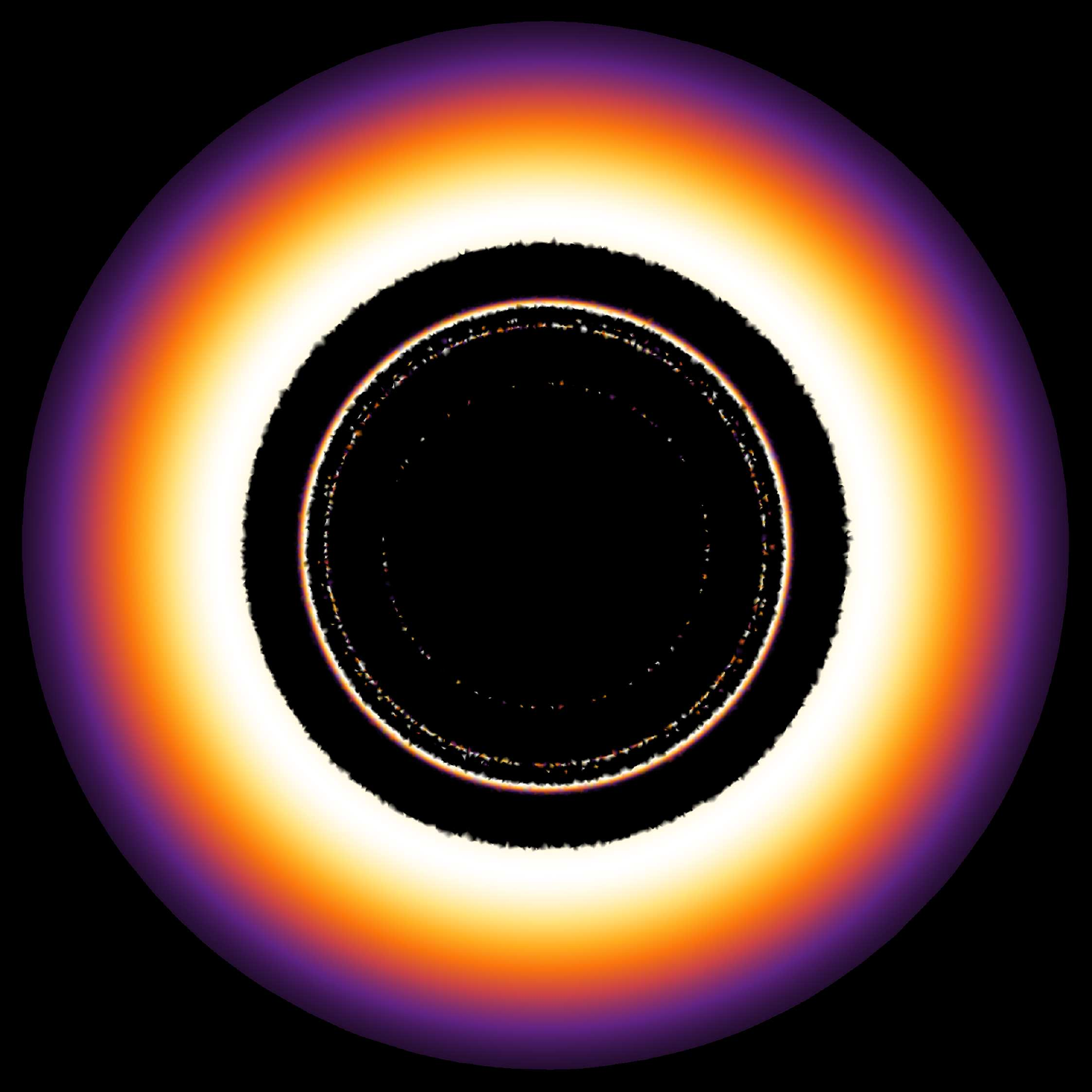}
  \hspace{0.5cm}
  \includegraphics[width=4cm,height=4cm]{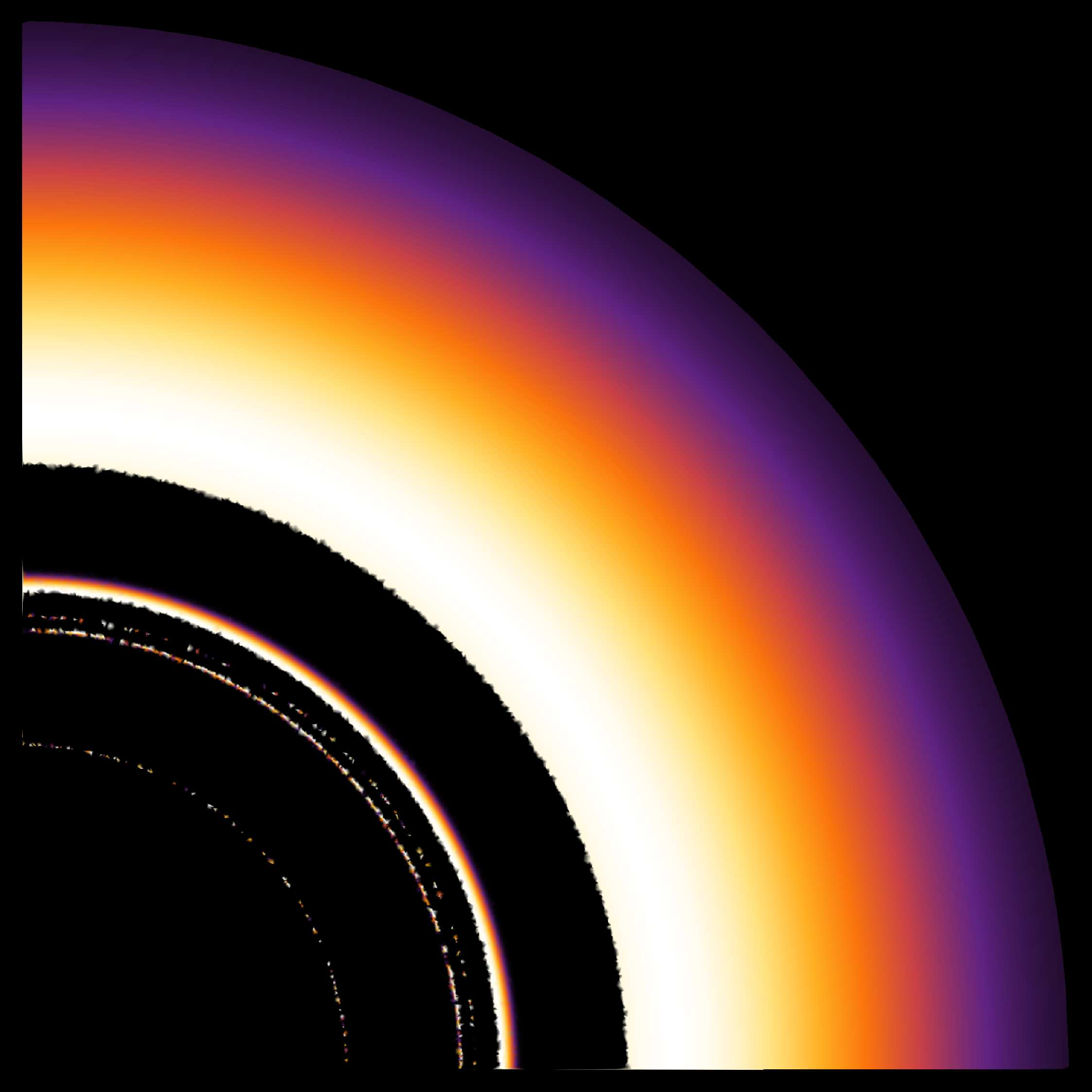}
  \hspace{0.5cm}
  \includegraphics[width=6cm,height=4cm]{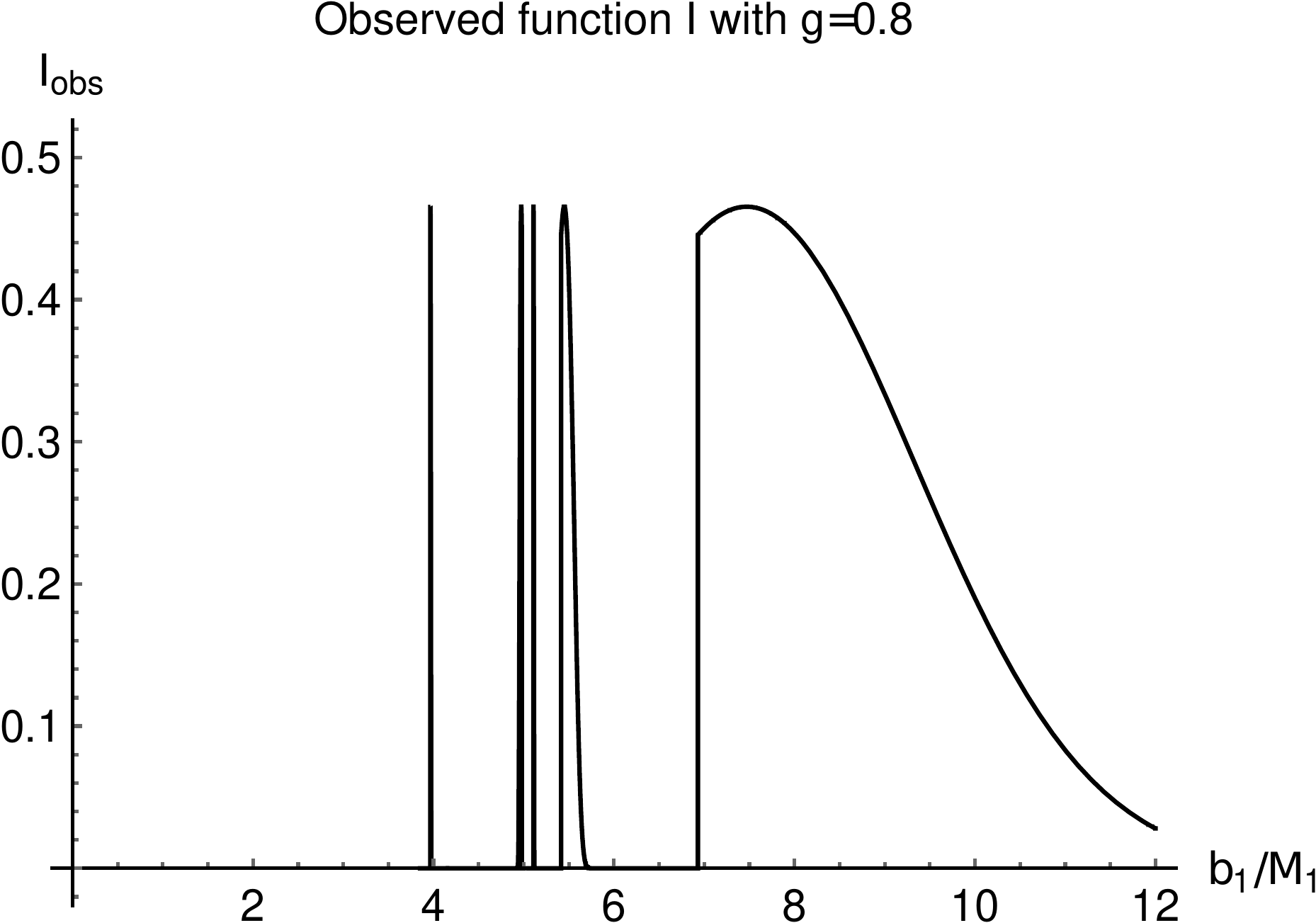}
  \hspace{0.5cm}
  \includegraphics[width=4cm,height=4cm]{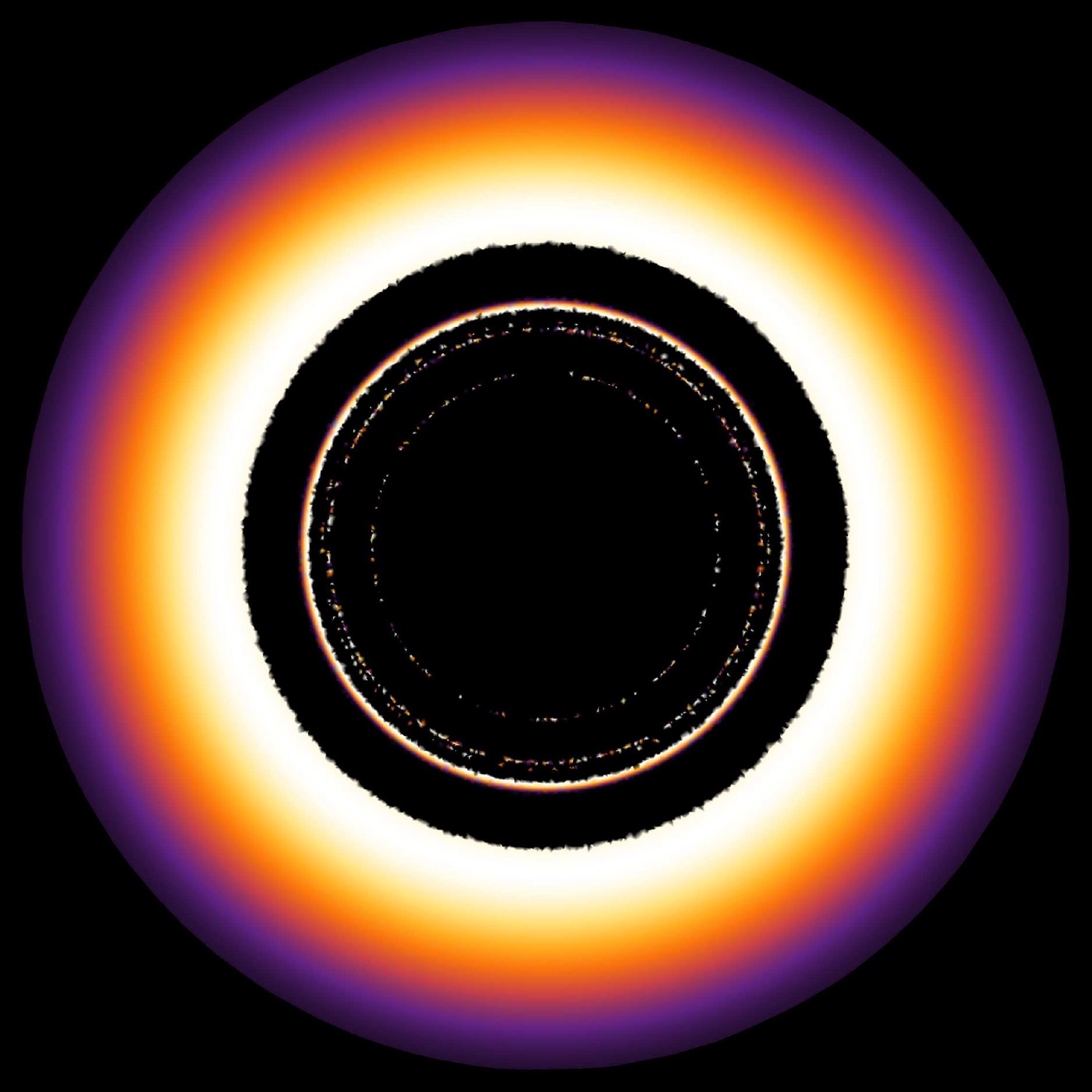}
  \hspace{0.5cm}
  \includegraphics[width=4cm,height=4cm]{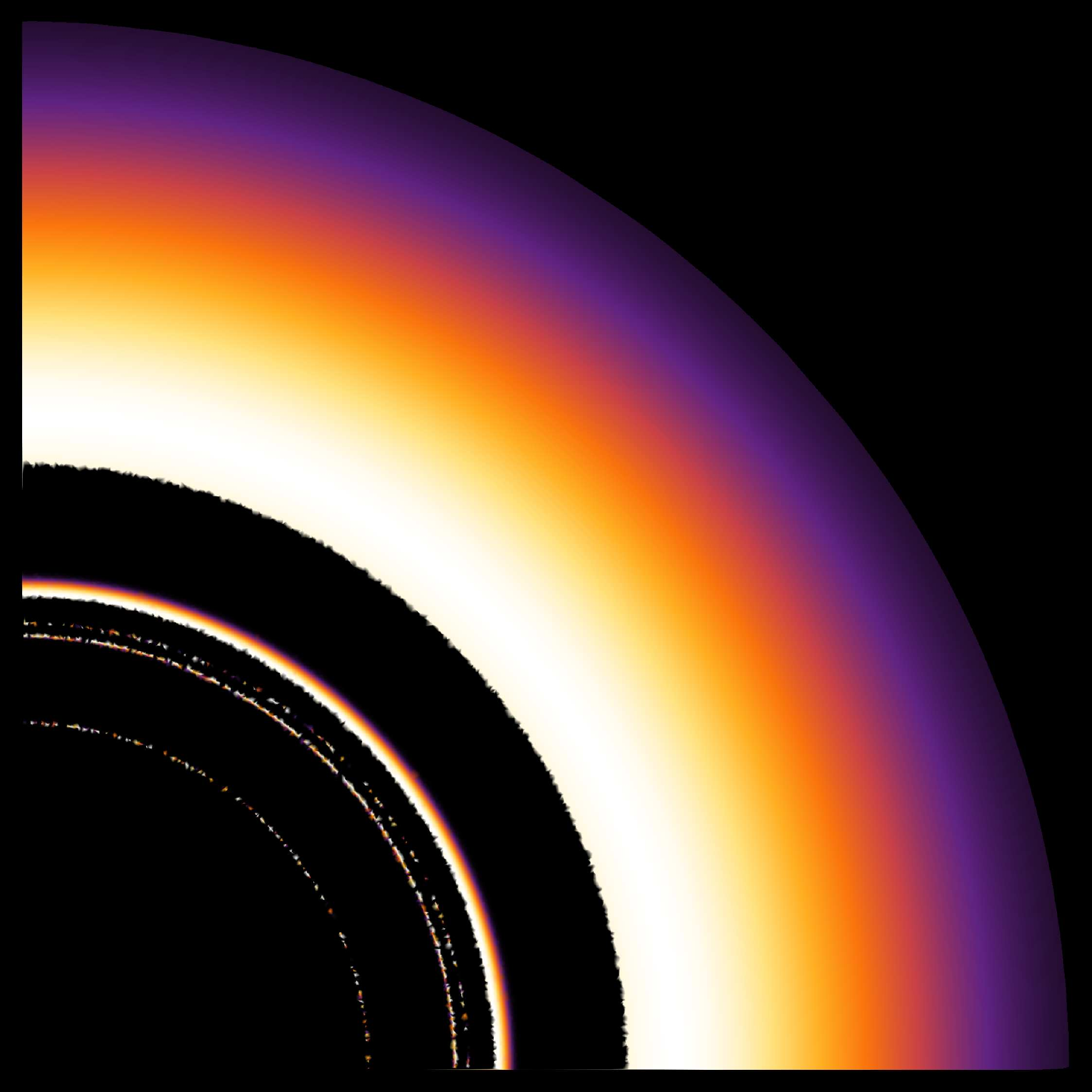}
\caption{\label{fig:7} \textbf{Model I}-- The total observed intensity as a function of the impact parameter ({\em left panel}) for the several representative magnetic charges and the corresponding two-dimensional images ({\em middle panel}) together with zooming in part of the image for illustrating their local density ({\em right panel}). The top, middle, and bottom panels are for $g=0$ (Schwarzschild TSW scenario), $g=0.5$, and $g=0.8$, respectively. The BHs mass are taken as $M_{1}=1$, $M_{2}=1.2$, and the throat radius is taken as $R=2.6$.}
\end{figure*}

\par
Figure \ref{fig:6} presents the radiation intensity as a function of the radius for both Model I and Model II. In Model I, we display the corresponding total observed intensity $I_{\rm obs}$ as a function of the impact parameter $b$, the two-dimensional image in celestial coordinates, and the local density image in Fig. \ref{fig:7}. The figure clearly illustrates the distinct regions of direct emission, lensing band, and photon ring group. Taking the case of magnetic charge $g=0.5$ (middle panel of Fig. \ref{fig:7}), we observe that the direct emission begins at around $b \simeq 7.04M_{1}$ and reaches its peak at approximately $b \simeq 7.56 M_{1}$. The maximum intensity is about $0.47$. The lensing band is confined to a narrow range of $b \simeq 5.57M_{1} \sim 5.83M_{1}$. The photon ring group appears at $b \simeq 5.17 M_{1}$, $b \simeq 4.93 M_{1}$, and $b \simeq 3.69M_{1}$. In the two-dimensional image, the boundary of the black disk corresponds to the starting position of direct emission. Within the black disk, a bright lensing band is visible, while the positions of the photon ring group progressively move towards the central region, appearing as several weaker rings. Notably, one of the new photon rings is located near $H b_{\rm c_{2}}$, and the other new photon ring is positioned inside the lensing band but outside $b_{\rm c_{1}}$. It is worth mentioning that the new second transfer function is not significantly affected in this model. The top and bottom panels of Fig. \ref{fig:7} depict the scenarios of a TSW with a Schwarzschild profile and a TSW with a Hayward profile, both under a magnetic charge of $g=0.8$. It is evident that an increase in the magnetic charge value causes the new photon ring to shrink inward towards the BH.

\par
\begin{figure*}[tbp]
\centering
  \includegraphics[width=6cm,height=4cm]{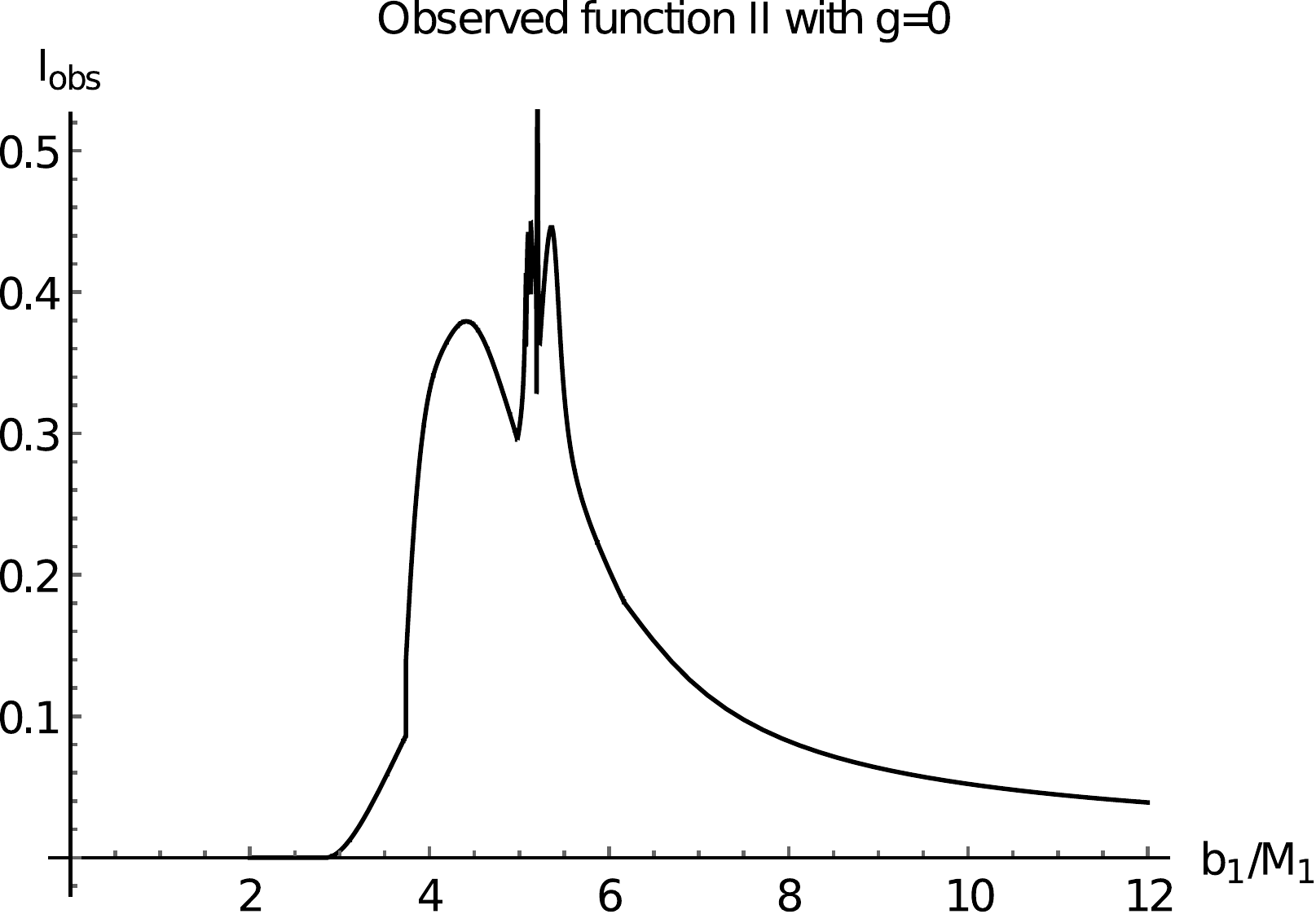}
  \hspace{0.2cm}
  \includegraphics[width=4cm,height=4cm]{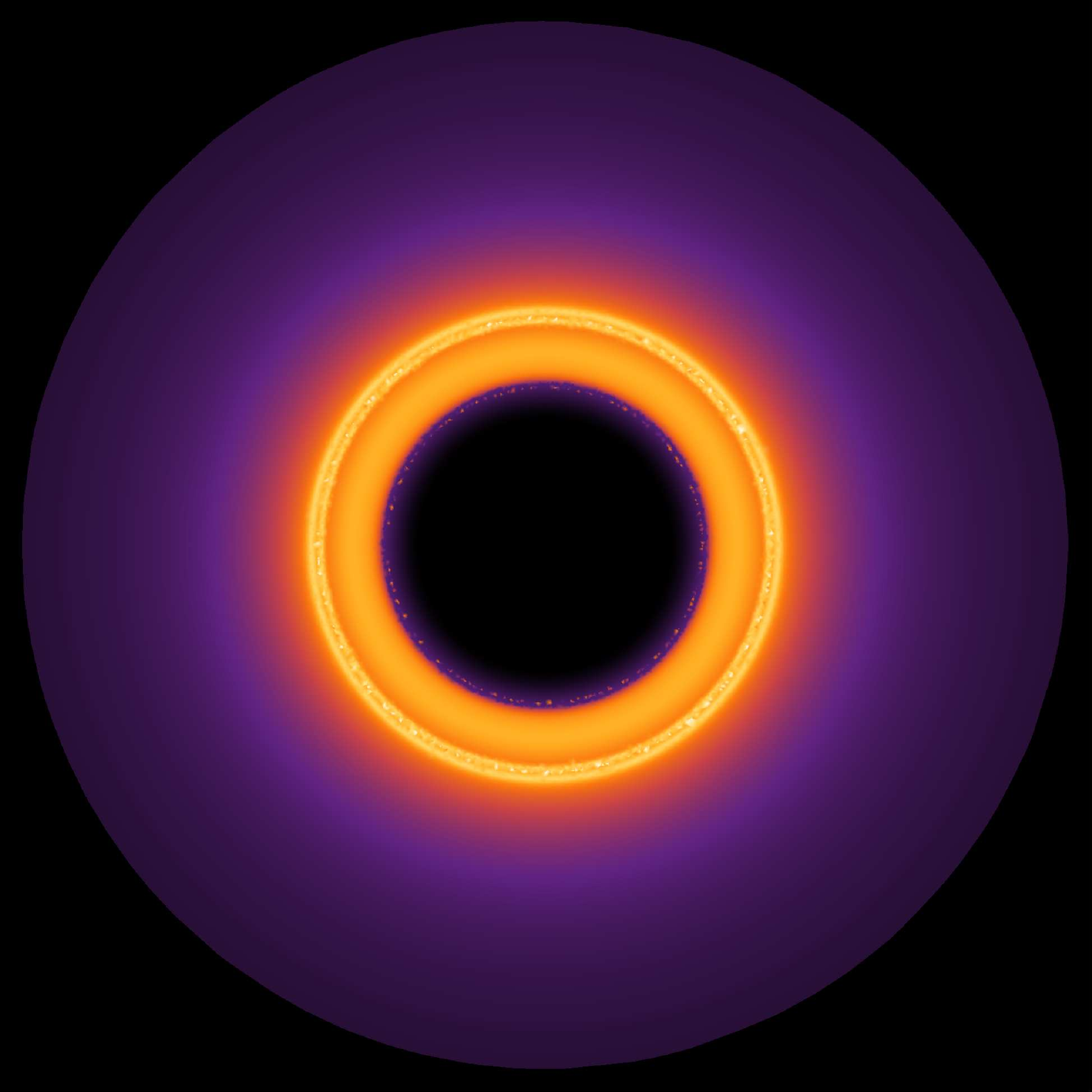}
  \hspace{0.5cm}
  \includegraphics[width=4cm,height=4cm]{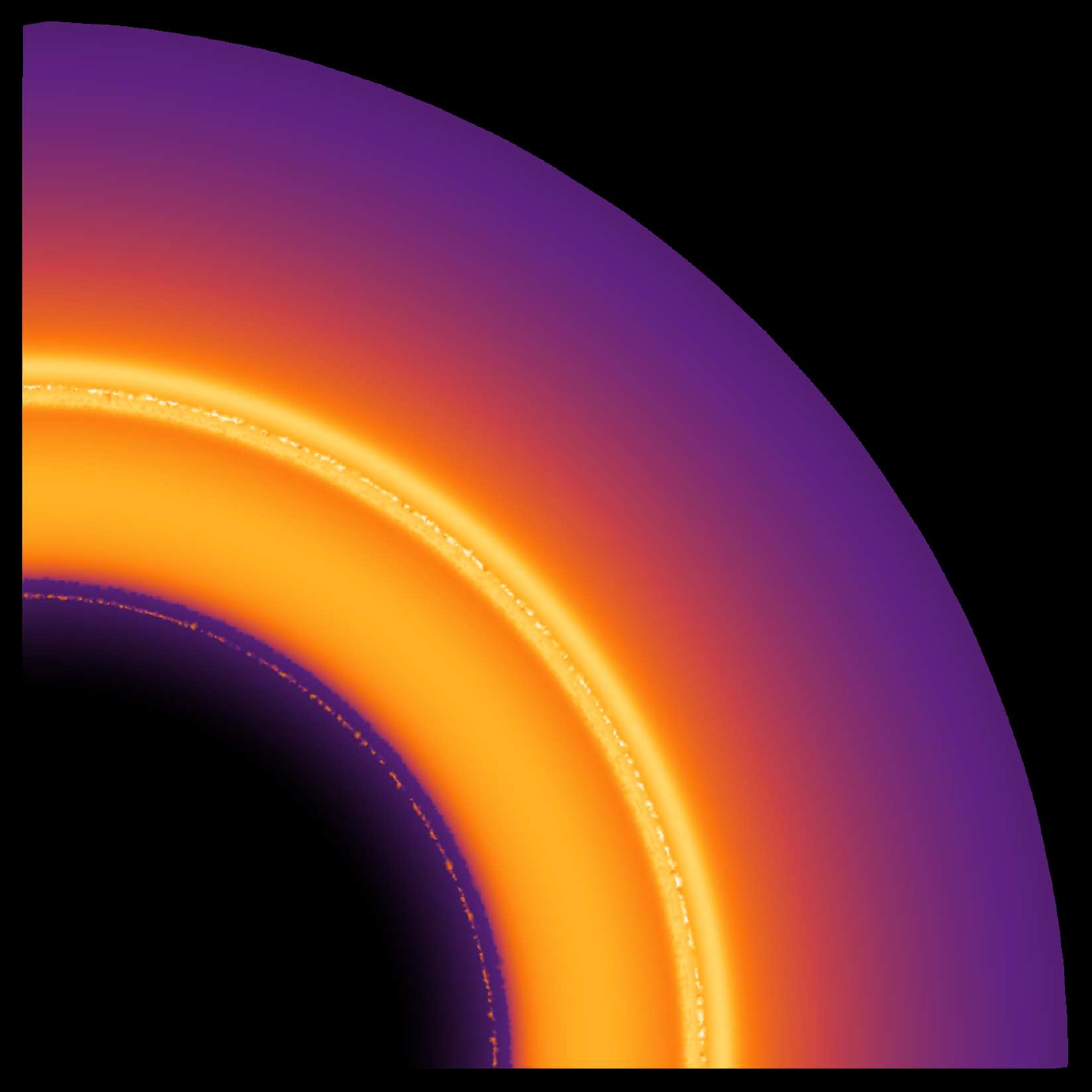}
  \hspace{0.5cm}
  \includegraphics[width=6cm,height=4cm]{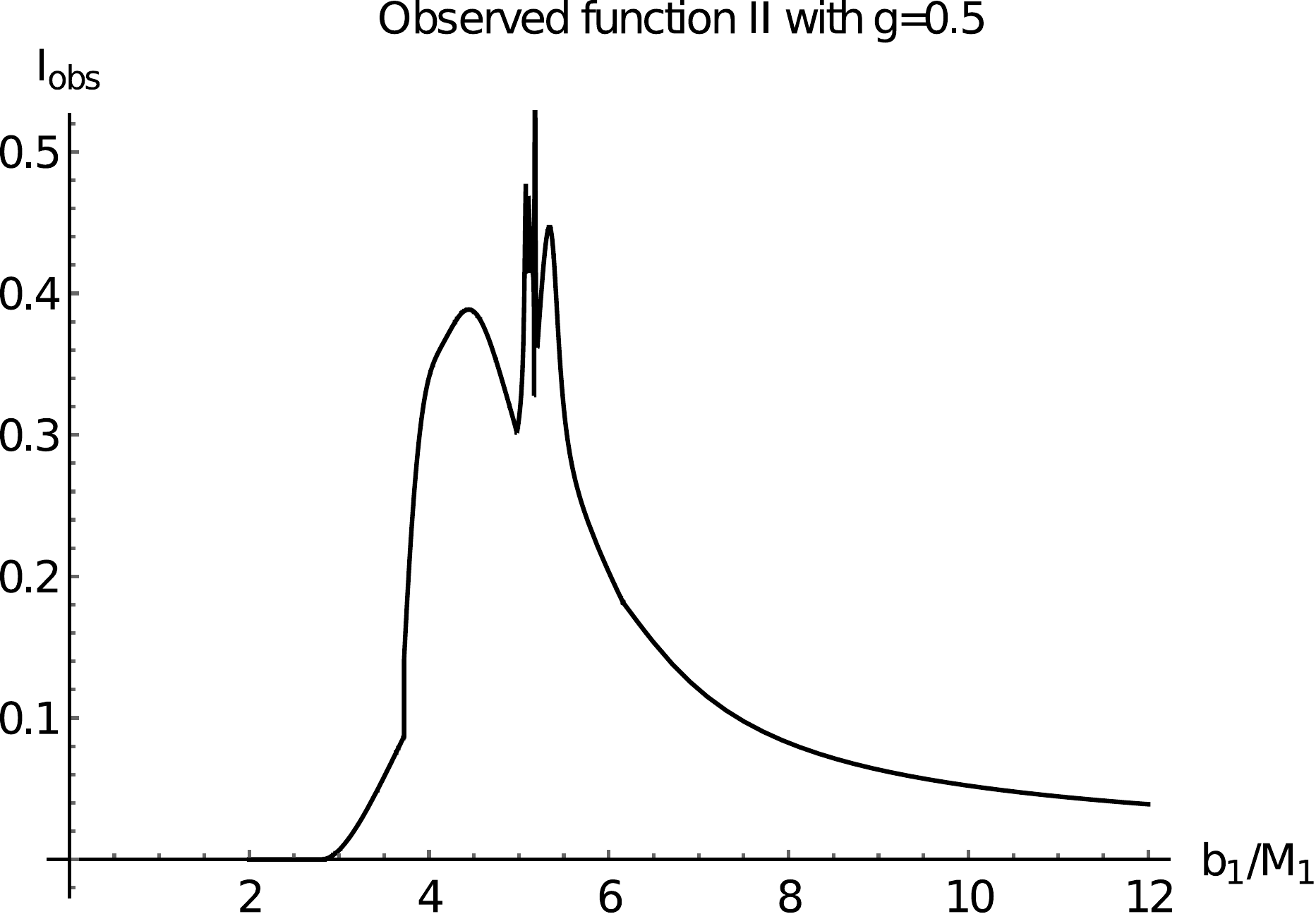}
  \hspace{0.2cm}
  \includegraphics[width=4cm,height=4cm]{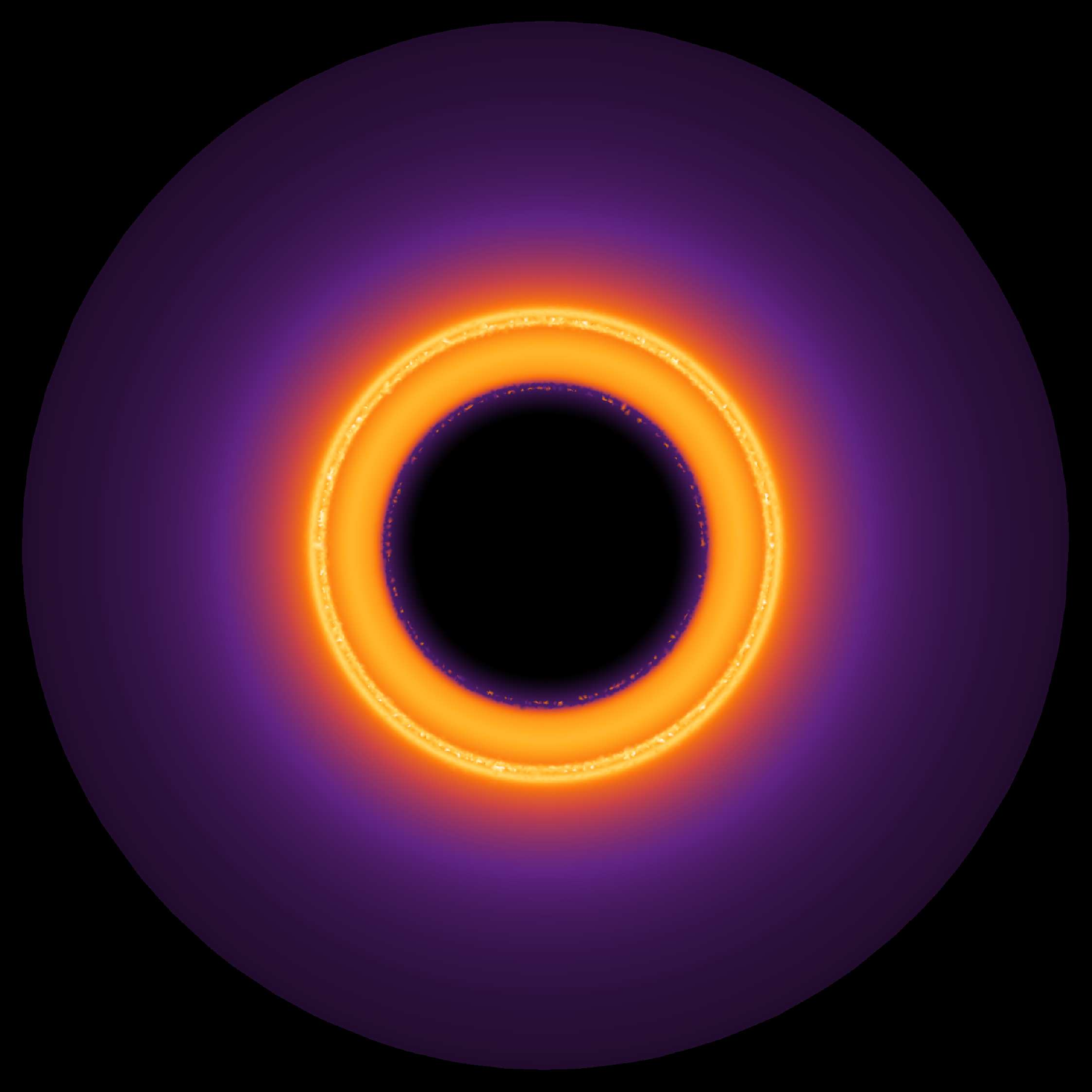}
  \hspace{0.5cm}
  \includegraphics[width=4cm,height=4cm]{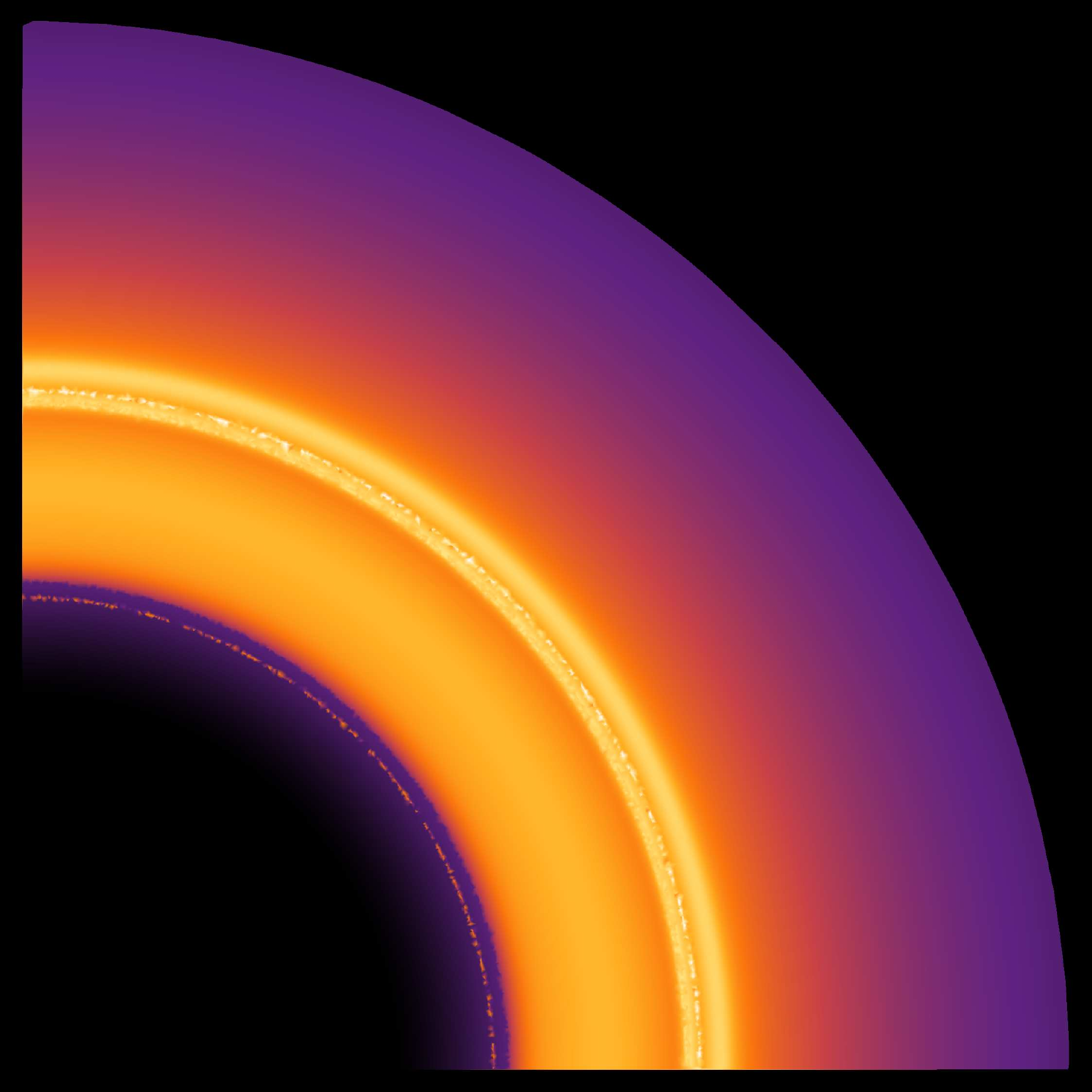}
  \hspace{0.5cm}
  \includegraphics[width=6cm,height=4cm]{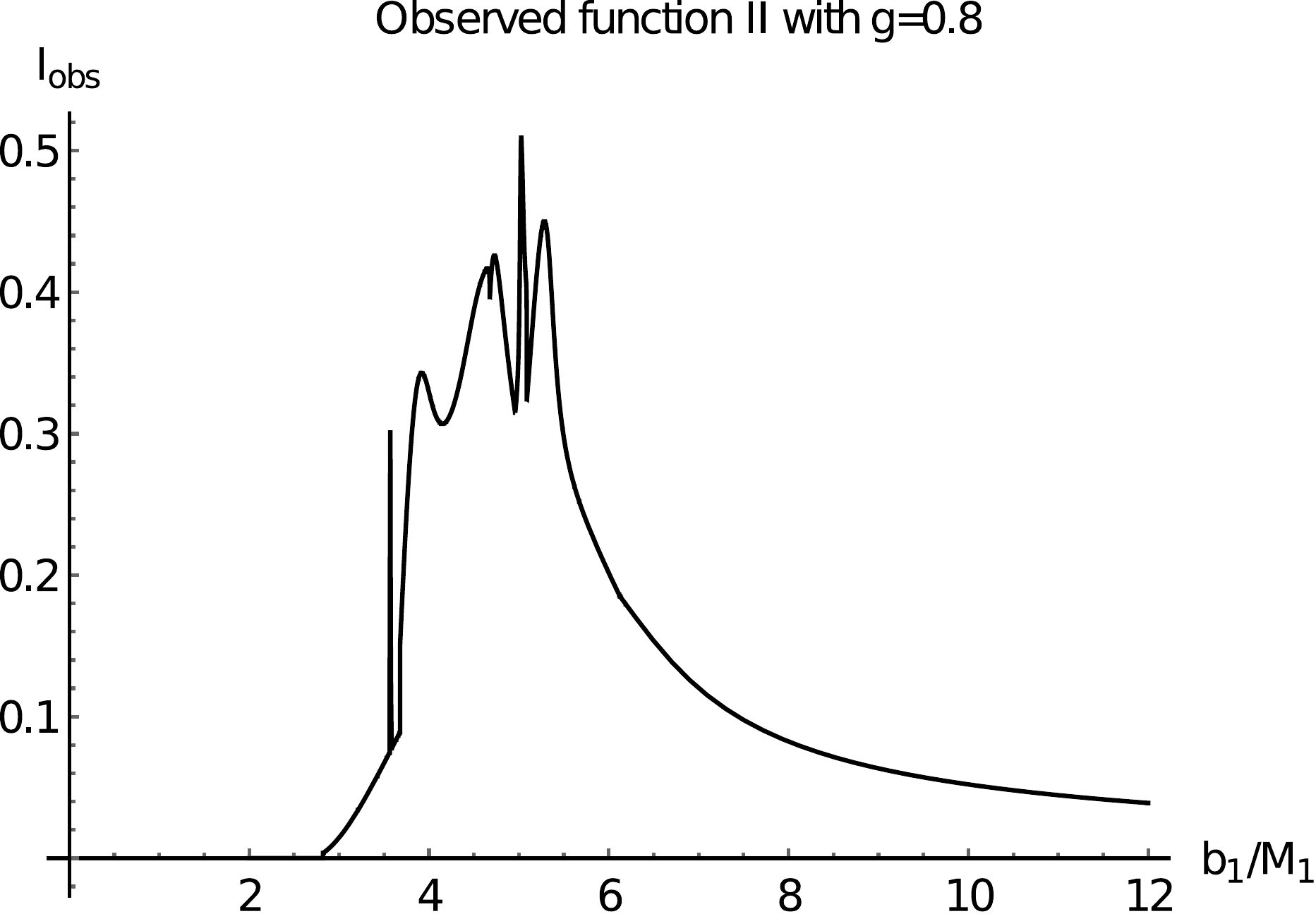}
  \hspace{0.2cm}
  \includegraphics[width=4cm,height=4cm]{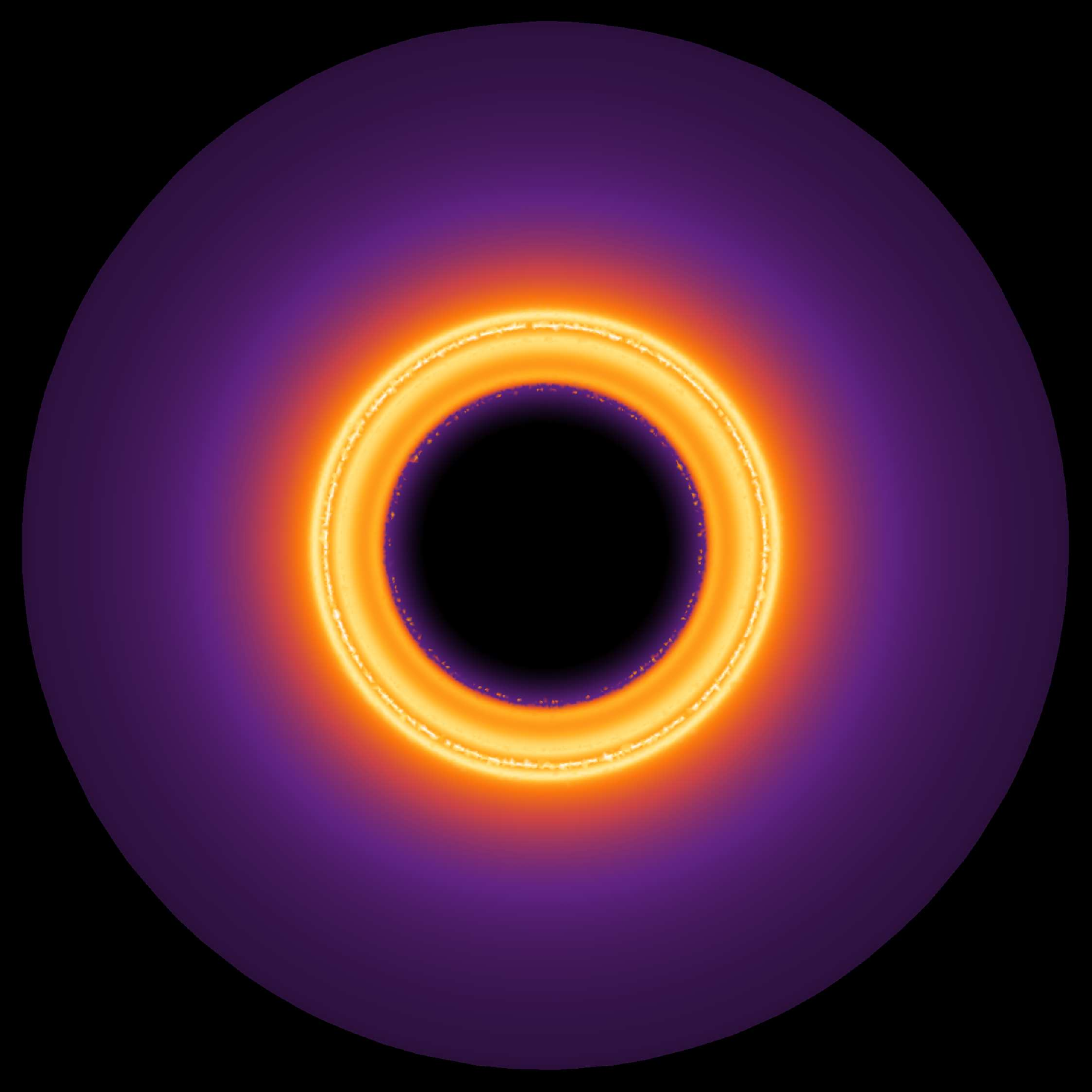}
  \hspace{0.5cm}
  \includegraphics[width=4cm,height=4cm]{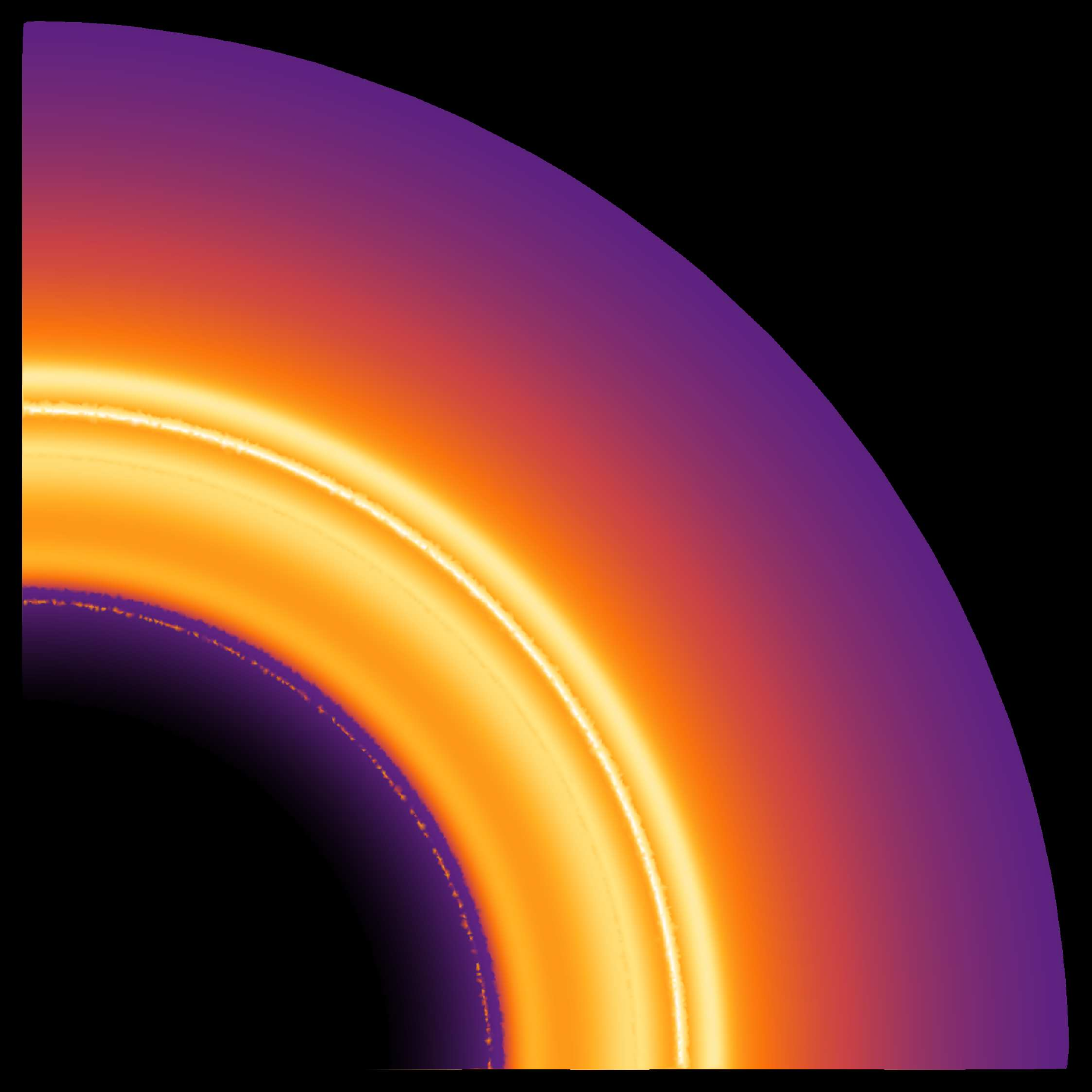}
\caption{\label{fig:8} \textbf{Model II}-- The total observed intensity as a function of the impact parameter ({\em left panel}) for the several representative magnetic charges and the corresponding two-dimensional images ({\em middle panel}) together with zooming in part of the image for illustrating their local density ({\em right panel}). The top, middle, and bottom panels are for $g=0$ (Schwarzschild TSW scenario), $g=0.5$, and $g=0.8$, respectively. The BHs mass are taken as $M_{1}=1$, $M_{2}=1.2$, and the throat radius is taken as $R=2.6$.}
\end{figure*}

Figure \ref{fig:8} showcases the total observed intensity as a function of the impact parameter, the two-dimensional image, and the local density image for Model II. In this case, we can observe that the regions of direct emission, lensing band, and photon ring overlap with each other. The photon ring group in Model II mentioned in the text overlaps with the region outside the photon sphere. Let's consider the case of a magnetic charge $g=0.5$ (middle panel of Fig. \ref{fig:8}). The direct emission begins at approximately $b \simeq 2.85M_{1}$, and the photon ring group is embedded within the lensing band, forming a distinct and bright multi-layered ring structure. Notably, the lensing band consists of the usual lensing ring and an additional lensing band. The usual lensing band is confined within the range of $b \simeq 5.24M_{1} \sim 5.51M_{1}$, while the additional lensing band is situated between $H b_{\rm c_{2}}$ and $b_{\rm c_{1}}$. This indicates that the new second transfer function contributes to the observed intensity in the case of a TSW with a Hayward profile in this radiation model.

\section{\textbf{Conclusions}}
\label{sec:5}
\par
In this analysis, we have extended the investigation of the BH shadow to the scenario of a TSW and explored the optical characteristics of a horizonless object with a Hayward profile. Through the construction of the TSW model, we obtained the effective potential of a TSW with a Hayward profile and plotted the potential function curve. The findings demonstrate that a throat connects two distinct spacetimes, and the effective potential function of the contralateral spacetime has the ability to reflect a significant portion of light back to its original spacetime.

\par
By utilizing the null geodesic equation, we have derived a specific expression for the radial component and determined the critical impact parameter for various magnetic charges. Our findings reveal that as the magnetic charge ($g$) increases, the critical impact parameter ($b_{\rm c_{i}}$) decreases, indicating that the photon ring moves closer to the BH. Moreover, we have examined the motion of photons in a TSW with a Hayward profile. It is observed that the deflection of light can be classified into three scenarios, depending on the mass relationship between the two spacetimes. When the impact parameter satisfies $H b_{\rm c_{\rm 2}} < b_{1} < b_{\rm c_{\rm 1}}$, the photons enter spacetime $\mathcal{M}_{1}$ from infinity, pass through the throat to spacetime $\mathcal{M}_{2}$, then return to spacetime $\mathcal{M}_{1}$ through the throat before finally exiting to infinity. Additionally, we have calculated the total deflection angle ($\phi$) of a TSW with a Hayward profile for photons with different impact parameters.

\par
By comparing the number of orbits between a BH and a TSW, we can analyze the behavior of photons with different total orbit numbers. Fig. \ref{fig:4} illustrates the variation of the total orbit number with respect to the impact parameter. It is evident that the orbit function $n_{1}$ of the TSW is equivalent to that of the BH, indicating that the image seen by an observer of the TSW contains the BH. Furthermore, the TSW situation introduces additional orbit functions $n_{2}$ and $n_{3}$, implying the presence of additional ring structures in the TSW image. To determine the observed intensity, we investigated its relationship with the radiation intensity and performed a ray-tracing procedure to obtain the transfer function $r_{\rm n}(b)$ for a TSW with a Hayward profile. Our results reveal that the first transfer function of a TSW with a Hayward profile closely resembles that of a Hayward BH. The second transfer function exhibits a monotonically increasing trend with an irregular segment preceding the monotonic part, indicating the existence of a ``lensing band'' in the resulting image. Additionally, in addition to the usual third transfer function of the BH, we identified two new third transfer functions in the TSW scenario, leading to the observer perceiving a ``photon ring group''.

\par
Building upon the previous discussions, we have conducted a comprehensive analysis of the optical appearance of a TSW with a Hayward profile. In order to characterize the radiation emitted by the accretion disk, we have parameterized its function using a Gaussian distribution. By considering two different Gaussian radiation models, we have investigated the observed function and the corresponding two-dimensional image. For Model I, we observed that the regions of direct emission, lensing band, and photon ring group are distinct and separated. The direct emission region starts at a specific impact parameter and reaches its peak intensity at another impact parameter. Within the black disk, a bright lensing band is clearly visible, while the photon ring group gradually moves closer to the central region, appearing as a series of weaker rings. In contrast, for Model II, the regions of direct emission, lensing band, and photon ring group overlap with each other. The photon ring group is embedded within the lensing band, forming a bright and distinctive multilayered ring structure. Notably, the lensing band consists of both the usual lensing ring and an additional lensing band. These findings highlight the potential of the optical appearance of a TSW to serve as a qualitative observational signature for the detection of wormholes in the future. By studying the distinct features and spatial arrangements of the direct emission, lensing band, and photon ring group, researchers can potentially identify and characterize TSW in astrophysical observations.

\setlength{\parindent}{0pt}\textbf{\textbf{Acknowledgments}}
This work is supported by the National Natural Science Foundation of China (Grant No. 12133003).\\

\end{CJK}
\end{document}